\documentclass[aps,physrev,groupedaddress,twocolumn]{revtex4-2}

\usepackage{float}
\usepackage{longtable}
\usepackage{graphicx}

\begin{document}

%\preprint{}

\title{Data evaluation processes of different-sized datasets: an eye-tracking with concurrent think-aloud study}


\author{Gregor Benz}
 \email{Contact author: gregorbenz@uni-koblenz.de}
 %\homepage[]{https://orcid.org/0000-0003-4506-2160}
\affiliation{%
 Physics department, University of Koblenz, Koblenz 56070, Germany
}%

\author{Sarah Taha}
\affiliation{
 Luitpold High School, Munich 81925, Germany
}%

\author{Andreas Vorholzer}
 %\homepage[]{https://orcid.org/0000-0002-0915-2516} 
\affiliation{%
 Physics department, University of Koblenz, Koblenz 56070, Germany\textsuperscript{}
}%

\date{\today}

\begin{abstract}
Considering the increasing availability of digital tools and measurement data in physics classrooms, students are more frequently confronted with larger datasets. While existing literature often assumes that working with large amounts of data is more complex, recent work by Benz, Ludwig, and Vorholzer [Sci. Ed. 119, 1669–1700 (2025)] suggests that this is not necessarily the case when data are presented in diagrams. Building on this, the present study investigates how students evaluate small and large datasets in diagrammatic representations. Using a process-oriented approach, we analyzed the data evaluation processes of $N=20$ university physics students via eye-tracking and concurrent think-aloud. The results indicate that dataset size, in interaction with the visibility of patterns in the data, systematically shapes how students reason with measurement data. Larger datasets support more pattern- and trend-based evaluation and can lead to more unambiguous conclusions. In contrast, smaller datasets are associated with a stronger focus on single measurement points and increased expressions of uncertainty, including an articulated ``need for data," alongside an overinterpretation/overweighting of single measurements. The observed shift from uncertainty-related and locally focused evaluation toward more integrative and trend/pattern-based reasoning suggests that larger datasets may support students in integrating multiple data points into more coherent interpretations of measurement data. Overall, the study contributes to a process-oriented understanding of how students engage with measurement data in physics education.
\end{abstract}

%\keywords{}

\maketitle

\section{Introduction\label{sec:Introduction}}

``Working with data is at the heart of science investigation"~\cite[p.\,280]{NationalAcademies}. Empirically based conclusions are commonly attributed a particularly high epistemic value, as they go beyond heuristic or purely theory-based considerations~\cite{Thornton1992,Volkwyn2005}. Accordingly, the importance of developing data-handling skills is reflected in various educational standards~\cite{NRC2012,NGSS2013,KMK2024,UK2014}.

Ongoing advances in the digitalization of measurement have led to a substantial increase in both the availability and the amount of data, not only in research but also in educational settings~\cite{Boyd2012,Duggan2002,Kjelvik2019}. In particular, the availability of digital measurement acquisition systems, which enable automated data collection and, thereby, allow students and teachers to collect large amounts of measurement data, makes large datasets an available resource for physics classrooms. While digital measurement acquisition systems have been available for physics classrooms for decades, their coherent integration into teaching and learning has yet to be achieved~\cite{Rosenberg2022,Benz2026a}.

A crucial issue is the potential benefits and drawbacks of using large datasets for educational purposes~\cite{Benz2024,Benz2025}. While it is generally agreed that conclusions derived from large amounts of measurement data can be considered more empirically sound than conclusions drawn from smaller amounts of measurement data, it remains unclear to what extent this stronger epistemic basis represents an added value for students’ data-based reasoning and conclusion drawing when learning about physical phenomena~\cite{Benz2025,Benz2022}. On the one hand, it seems plausible that larger amounts of measurement data enable more unambiguous conclusions about physical phenomena than smaller amounts, as they provide more precise evidence. On the other hand, large amounts of measurement data can also lead to conclusions that are less straightforward to draw from the students’ perspective, particularly when multiple measurands influence one another or when longer or faster measurements reveal new or different patterns in the measurement data.

The benefits and drawbacks of large datasets are especially relevant in cases where the patterns that can be observed in the data (and the conclusions these patterns support) depend on the size of the dataset. This is the case, for instance, when a physical phenomenon can only be observed with a particularly high sample rate or when data is collected over a long time interval~\cite{Benz2022}. Understanding how students engage with and interpret measurement data when larger amounts of measurement data reveal new or different patterns, which data elements they attend to, and how they select relevant information when drawing conclusions is, therefore, an important step towards using large datasets in physics classrooms more effectively.

Existing research on students’ engagement with measurement data has predominantly focused on outcome-level measures and correlative relationships, thereby offering limited insight into the perceptual and cognitive processes during data evaluation in experimental contexts. As a result, the strategies by which students make sense of different amounts of measurement data remain largely unexplored. Addressing this gap requires a process-oriented perspective on students’ data evaluation, focusing on how information is visually attended to and processed during data evaluation. The study presented here pursues this approach by capturing and comparing the data evaluation processes of students engaging with different amounts of measurement data using eye-tracking (with concurrent think-aloud).

We begin by providing relevant background information about how students deal with data in physics classrooms (Section~\ref{sec:Background}) and derive our research question from this (Section~\ref{sec:Research question}). This is followed by a description of the methodology used in our investigation (Section~\ref{sec:Methods}). The results (Section~\ref{sec:Results}) and discussion (Section~\ref{sec:Discussion}) present the findings of the study and their interpretation. Finally, the conclusion (Section~\ref{sec:Conclusions}) summarizes the key findings and their implications for instruction and future work.

\section{Background\label{sec:Background}}

\subsection{Consequences of dealing with large amounts of measurement data compared to small ones\label{sec:Consequences of dealing with large amounts of measurement data compared to small ones}}

Existing research has already investigated how students deal with measurement data in physics education, including effects on students’ reasoning or dealing with measurement uncertainties. Beyond dealing with measurement data in general, the literature indicates that handling different amounts can affect how it needs to be handled. First, datasets of different sizes can differ in their nature and management. Small datasets (often characterized by fewer than 40 data points and two or fewer variables acquired in parallel~\cite{Benz2025,Benz2024,Rosenberg2022}) are typically described as structured and allowing for intuitive exploration of physical phenomena using paper-and-pencil tools~\cite{Benz2025,Benz2024,Kastens2015,Kjelvik2019}. This promotes direct, exploratory practices without technical difficulties. In contrast, large datasets are often described as being more heterogeneous and containing outliers. This necessitates strategic management, advanced analytical tools, data wrangling for cleaning, and digital tools (e.g., Excel, R) for handling~\cite{Schultheis2015,Benz2025,Fielding2025,Kastens2015,Kjelvik2019,Rosenberg2022,Wilkerson2025,Lee2018}. Second, the amount of data can impact the epistemic uncertainty. When collecting small datasets first-hand, sources of error are often tangible and can be reflected upon~\cite{Kastens2015,Lee2018}. However, these differences are not solely attributable to dataset size, but also to the context of data collection and representation, and are an attribute of first-hand data collection. This enables transparent evaluation of data trust. In contrast, large second-hand datasets often require critical reflection due to repurposing and limited metadata, which can lead to spurious correlations~\cite{Kastens2015,Lee2018,Fielding2025,Wilkerson2025}, highlighting that differences attributed to dataset size may in practice co-occur with differences in data origin and structure. This requires meta-knowledge-based validation and sensitivity to experimental contextual errors. Third, the size of the dataset may impact the knowledge that can be gained. Small datasets can provide focused insights into processes or laws~\cite{Kastens2015,Kjelvik2019,Rosenberg2022,Lee2018} and thereby support the conceptual understanding of individual phenomena. Large datasets, in contrast, can extend this conceptual understanding by enabling access to multivariate patterns that become visible only at higher numbers of data points. These patterns may require iterative filtering strategies and can reveal structures that go beyond those observable in small datasets~\cite{Kastens2015,Resnick2018,Wilkerson2025,Benz2022}. As a consequence, large datasets enable a broader and more statistically grounded form of pattern recognition, albeit at the cost of increased analytical complexity. Lastly, different amounts of measurement data can convey different epistemological images of the nature of science. Small datasets can convey a more linear and goal-oriented image of science with ``correct" results~\cite{Bowen2007,Lee2018,Schang2023}. In contrast, large amounts of data can convey an image of science as an iterative and uncertain processes that require, among other things, the consideration of bias~\cite{Schultheis2015,Lee2018,Bowen2007,Fielding2025}. We want to emphasize that whether the four identified differences between using small and large datasets occur in practice depends on the specific scenario (e.g., how data is presented and discussed). Not all large datasets automatically convey a more accurate image of science and small datasets may very well require critical reflection. Rather, the potential differences represent analytically distinguishable categories that repeatedly emerge in the literature on data practices and physics learning. Together, these aspects provide a heuristic for understanding how increasing data amounts can impact data evaluation processes.

In addition to the \textit{amount} of data itself, data evaluation processes may also be impacted by how patterns and structures become visible in \textit{the data representation}. For instance, when it comes to the analysis of time-series data, periodic or quadratic trends may only emerge clearly with higher sampling rates or longer observation intervals, while smaller datasets may obscure such patterns. When analyzing students data evaluation processes, it is often difficult to disentangle the effects of data size from the effects of data representation. It is, therefore, important to consider both effects jointly when analyzing such processes. It can be assumed that the amount of data as well as the data representation do not inherently facilitate or hinder data evaluation processes, but rather modulate existing difficulties and affordances. For instance, a larger number of data points in a representation can support trend or pattern recognition when additional measurements make underlying regularities more visible by reinforcing existing trends~\cite{Benz2025}. At the same time, larger datasets can increase the apparent variability and complexity of the data, potentially heightening students’ uncertainty during data evaluation~\cite{Masnick2008}. As a result, these modulated difficulties can lead to differences in data evaluation processes and task performance. For instance, simpler trend/pattern recognition can steer students toward big-picture analysis rather than selective, local interpretations of single measurement data points~\cite{Boaventura2013,Kuhn1995,Masnick2008,Kanari2004}. Greater uncertainty can prompt heavier reliance on heuristics~\cite{Baur2018,GarciaMila2017,Ludwig2021,Lenz2025}, ignoring anomalous data~\cite{Baur2018,Greenhoot2004,Valanides2014} or contradictory evidence to the original (intuitive) hypothesis~\cite{Boaventura2013,Baur2018,Dunbar1993,Greenhoot2004,Hammann2008,Masnick2008,Park2006}, possibly via peripheral information processing~\cite{Ludwig2017}.

Existing evidence on how students handle different amounts of measurement data is mixed. For instance, a randomized study involving over 650 high-school students found that participants selected more correct conclusions from largest datasets (Bayesian Cohens’s $d=1.09$, 95\,\% CI\,[0.83, 1.33]), without perceiving data evaluation as more complex, despite increased data volume or changed pattern in the datasets~\cite{Benz2024,Benz2025}. At the same time, justifications for the conclusion selected in terms of content and structure remained unchanged~\cite{Benz2025}. Together, these findings indicate that simply examining outcomes of data evaluation is insufficient to understand how students evaluate different amounts of measurement data. This motivates a process-oriented investigation into the underlying visual and cognitive mechanisms of data evaluation, particularly as dataset size varies.

\subsection{Status quo on students' data evaluation processes and desiderata\label{sec:Status quo on students' data evaluation processes and desiderata}}

To better understand how students evaluate datasets of varying sizes, it is necessary to examine the underlying visual and cognitive processes involved in drawing conclusions from measurement data.

Studies of students' data evaluation processes consistently show that drawing conclusions from graphical data representations is closely linked to visual perception and attention~\cite{Hahn2022,Becker2023}. Building on the \textit{eye-mind hypothesis}~\cite{Carpenter1989}, which posits that the information units currently being fixated are cognitively processed, eye-tracking studies provide particularly detailed insights into perception-related and cognitive aspects of data evaluation. Numerous studies have shown that successful problem solvers focus their visual attention specifically on diagnostically relevant areas of diagrams, while less successful students often fixate on visually salient features that are irrelevant to the question at hand~\cite{Becker2023,Hahn2022,Madsen2012,Madsen2013}. These differences in gaze patterns are directly reflected in the conclusions drawn and illustrate the central role of visual attention in data-based reasoning.

Although eye-tracking studies provide detailed insights into visual and cognitive aspects of data analysis, their empirical focus has been relatively narrow to date. While some research focuses on different forms of data representation (overview in~\cite{Ruf2023}), another large part of eye-tracking research investigates data evaluation processes using idealized curves in diagrams~\cite{Brückner2020,Brueckner2020,Kekule2014,Klein2019,Klein2020,Klein2021,Madsen2012,Madsen2013,Rouinfar2014,Skrabankova2020,Susac2018}. These idealized curves in diagrams can be seen to clearly demonstrate relevant features, and there is little epistemic uncertainty. Against the backdrop of the contrasts outlined in Section~\ref{sec:Consequences of dealing with large amounts of measurement data compared to small ones}, such representations can also be characterized as small datasets, even if they differ fundamentally in their nature and origin. In such investigation settings, data evaluation is implicitly framed as a process of correctly extracting meaning from the given measurement data, with the primary goal of identifying the ``correct" interpretation~\cite{Hahn2022}. To the best of our knowledge, and in line with a recent review~\cite{Hahn2022}, we are not aware of any eye-tracking studies that involve data evaluation processes using diagrams that display measurement data points or different quantities of measurement data points. However, this focus leaves open how students evaluate data in less idealized settings, where measurement data are represented as discrete points and epistemic uncertainty is an inherent feature of the task. In such contexts, aspects such as selecting or reducing measurement data, handling competing interpretations, and critically examining uncertainty become central to the data evaluation process.

While eye-tracking has become a prominent approach for investigating students’ data evaluation processes, the \textit{eye-mind hypothesis} has certain limitations for reconstructing these processes~\cite{Kok2016}. On the one hand, a student may focus on a relevant feature in a diagram but interpret it incorrectly (e.g., the person identifies an increase in a body’s position data but interprets it as a standstill), or the person focuses on heuristics or prior knowledge. On the other hand, internal deliberations (e.g., evaluating competing interpretations, reflecting on epistemic uncertainty, or selectively weighing data points) remain also inaccessible through eye-tracking. These examples illustrate that although eye-tracking can provide indications of ongoing data evaluation processes, it does not allow for comprehensive interpretative conclusions about the corresponding thought processes. Approaches proposed in the literature to address these limitations include combining eye-tracking with think-aloud or multiple-choice answers with written explanations~\cite{Hahn2022,Kok2016}. However, such combined approaches remain relatively rarely implemented. To the best of our knowledge, only one study has examined data evaluation using such combined methods~\cite{Klassen2026}. This study investigated differences between experts’ and novices’ data evaluation processes when interpreting idealized graphs and showed that combining eye-tracking data with think-aloud protocols provides a more comprehensive account of data evaluation processes. However, studies investigating how students evaluate datasets that differ in size are, to our knowledge, still lacking.

\section{Research question\label{sec:Research question}}

While eye-tracking research has substantially advanced our understanding of visual attention in data evaluation, its application to realistic measurement data, characterized by discrete points, varying quantities, and inherent epistemic uncertainty, remains limited. Existing studies predominantly examine idealized curve representations that presuppose clear ``correct" interpretations. Although initial work has begun to combine eye-tracking with additional process measures, these investigations also focus on idealized representations and do not address how students evaluate diagrams presenting different amounts of measurement data. Moreover, it remains unclear which cognitive processes occur during fixation on diagram elements. These limitations are particularly relevant for physics education, where large amounts of measurement data (as opposed to small amounts) reveal new phenomena but require careful interpretation to draw appropriate conclusions. Therefore, we intend to answer the following research question:

How do students evaluate different amounts of data presented in a diagram?

\section{Methods \label{sec:Methods}}

This study analyzes the data evaluation processes of, $N=20$ university physics students (65\,\% male, 35\,\% female, average age 22.5 years). Eight of the 20 students were studying in a physics teacher program, while the remaining students were studying physics. At the time of data collection, the students were at different stages of their studies: five were freshmen, 7 others were also in a bachelor’s program, and 8 were in a master’s program.

\subsection{Introduction to the experiment \label{sec:Introduction to the experiment}}

\begin{figure}
 \includegraphics{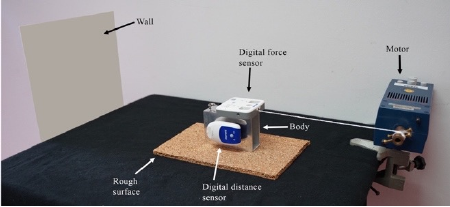}%
 \caption{Experimental setup as shown to the students~\cite{Benz2025}.\label{fig:Experimental setup as shown to the students.}}
 \end{figure}

At the beginning of the study, the students were introduced to the experimental setup and the collection of measurement data. In the experiment, a body was accelerated over a rough surface by a motor that rolled up a string connected to the body (see Figure~\ref{fig:Experimental setup as shown to the students.}). A digital force sensor acquired the acting force on the body, and a digital distance sensor acquired the body’s position on a wall.

 The experimental setup was presented to the students using a textual description and a picture of the setup (see Figure~\ref{fig:Experimental setup as shown to the students.}). If anything about the setup was unclear to the students, they could ask the experiment supervisors questions at any time. The experimental setup and context were successfully used in subsequent studies on students’ data interpretation~\cite{Benz2025}.

\subsection{Presentation and evaluation of measurement data \label{sec:Presentation and evaluation of measurement data}}

\begin{figure*}
 \includegraphics[width=\textwidth]{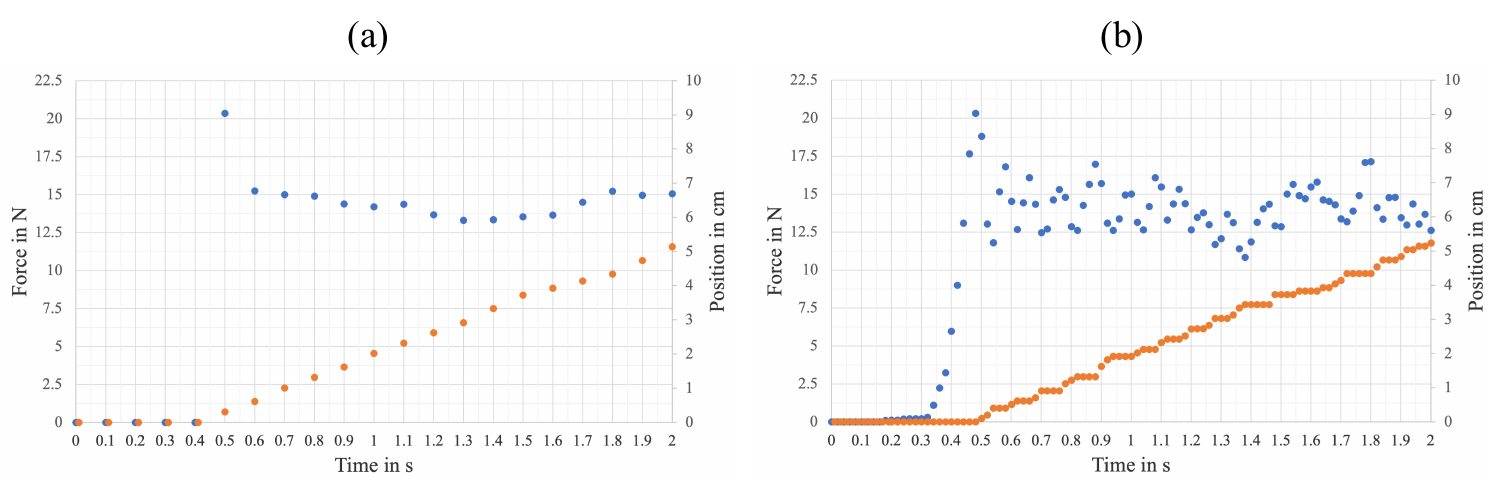}%
 \caption{Diagrams representing the (a) small or (b) large dataset~\cite{Benz2025}.\label{fig:Diagrams representing the (a) small or (b) large dataset.}}
 \end{figure*}

After an introduction to the experiment, the students were randomly assigned to one of two groups: Group S was first asked to evaluate the small dataset collected with a low sampling rate of 10 samples per second (Figure~\ref{fig:Diagrams representing the (a) small or (b) large dataset.}a) and then the large dataset collected with a high sampling rate of 50 samples per second (Figure~\ref{fig:Diagrams representing the (a) small or (b) large dataset.}b). Group L was asked to evaluate the datasets in reverse order. Although the order of presentation was reversed, the sequential evaluation of two datasets may still introduce carry-over effects, such as sensemaking about the underlying phenomenon or transfer of evaluation strategies between conditions.

%\begin{widetext}
 %\end{widetext}

 Importantly, in the present context, the manipulation of the sampling rate did not only affect the dataset size but also fundamentally changed the structure and visibility of patterns in the data. For the lower sampling rate (Figure~\ref{fig:Diagrams representing the (a) small or (b) large dataset.}a), the position data increase relatively uniformly over time, while the force data remain approximately constant during this increase. The data can be interpreted as the body being exclusively in a sliding phase with a relatively constant velocity. For the higher sampling rate (Figure~\ref{fig:Diagrams representing the (a) small or (b) large dataset.}b), the position data show alternating phases in which they increase uniformly and phases in which they remain constant, with these phases alternating rapidly. During phases of increasing position, the force data decrease, whereas during phases of constant position, the force data decrease, whereas during phases of constant position, the force data increase. The data can be interpreted as a repeated transition between sticking and sliding phases. As a result, dataset size and pattern visibility are confounded in the present design and cannot be disentangled.

For the evaluation of the dataset, participants in both groups first received a task that asked them to decide which phase was most likely present within the time interval from 1.7\,s to 1.9\,s: a sticking phase (Conclusion~1), a sliding phase (Conclusion2), or both phases (Conclusion~3) within the time interval of 1.7\,s to 1.9\,s. The conclusion was captured using a multiple-choice single-select task. The participants were then asked to provide a written justification for their conclusion. From a physics expert's perspective, Conclusion~2 is supported by the small dataset, whereas Conclusion~3 is supported by the large dataset.

\subsection{Capturing the data evaluation process \label{sec:Capturing the data evaluation process}}

Since the aim of the study was to examine the data evaluation processes that led the participants to the selected conclusion and written justification, we captured the data evaluation process via eye-tracking and concurrent think-aloud. Whereas eye-tracking allows for capturing which elements in the diagrams the participants considered when drawing their conclusions and justifying them, concurrent think-aloud provides deeper insights into the cognitive processes at work~\cite{Kiili2014,Becker2023}. A drawback of concurrent think-aloud is, however, that performing think-aloud may impact gaze movements~\cite{Hahn2022,Ibrahim2021} and, therefore, interferes with the eye-tracking data. To address this issue, we investigated one of the two data evaluation processes of each group using eye-tracking with concurrent think-aloud, and the other using ``pure" eye-tracking. In sum, these considerations led to four experimental conditions: S-ET, S-ET-TA, L-ET, L-ET-TA. The naming of the conditions is based on the size of the first dataset the participants in that group received (S\,$=$\,small; L\,$=$\,large) and the method used to capture the data evaluation process (ET\,$=$\,eye-tracking, ET-TA\,$=$\,eye-tracking with concurrent think-aloud) for the first dataset.

To evaluate the diagrams, participants sat at a 22-inch computer screen with a resolution of 1680\,x\,1050 pixels. The distance between the participants and the screen was about 70\,cm. To capture gaze behavior during eye-tracking (with concurrent think-aloud), we used a Tobii Pro Fusion eye tracker with a sampling frequency of 250\,Hz and an accuracy of 0.30° of visual angle (according to the manufacturer)~\cite{TobiSoftware2024}. Since we were interested in which elements the participants evaluated in the dataset, fixations were detected using an I-VT algorithm~\cite{Salvucci2000}.

\subsection{Data analysis \label{sec:Data analysis}}

As part of the data evaluation, we captured three types of data evaluation process data: (1) the written justifications for/against the selected conclusions, which can be understood as a summary of the elements considered (most) relevant in the diagrams and around the experiment, and which were written down after the data evaluation, (2) the eye-tracking fixation data, which provides information about which elements in the diagrams the participants considered during the data evaluation, and (3) the thoughts expressed during the data evaluation, which provide information about elements in the diagrams and around the experiment that were considered.

To identify which elements the participants considered during data evaluation, we have inductively developed two coding manuals: one for the written justifications and the concurrent think-aloud data (see Appendix~\ref{sec:Coding manual for written justifications and think-aloud data}) and one for the eye-tracking fixation data (see Appendix~\ref{sec:Coding manual for eye-tracking fixation data}). The procedures for developing both manuals were identical, with the only difference being the underlying data (written justifications and concurrent think-aloud data, or eye-tracking fixations). We started by identifying elements that the participants referred to in the written justifications, during the concurrent think-aloud, and fixated on in the diagram. We then defined these elements as codes and phrased rules for when they apply. Next, we double-coded all written justifications, concurrent think-aloud transcripts, and scan plots that showed fixations. Interrater agreement was calculated and can be considered as `substantial' to `almost perfect' for all types of data evaluation process data (written justification: Cohen’s $\kappa = 0.75$, eye-tracking data: Cohen’s $\kappa = 0.81$; concurrent think-aloud data: Cohen’s $\kappa = 0.80$)~\cite{Landis1977}. Then, we discussed the discrepancies until we reached agreement on the coding results (Cohen’s $\kappa = 1.00$ for all types of data). Due to the relatively small database, coding was not carried out in several rounds to iteratively increase interrater reliability; however, the coding rules were refined during discussion to reach agreement.

To answer the research question, we sorted the elements referred to/fixated on by the amount of measurement data evaluated (groups S and L) and by the conclusion selected within this grouping. We did the latter against the background that it can be assumed that the elements considered, as well as the strategies pursued regarding the data evaluation process, differ to reach a specific conclusion.

\section{Results \label{sec:Results}}

When the small dataset was evaluated, in 75\,\% of the evaluations, the participants concluded that the body is in both motion states during the corresponding time interval (Conclusion~3), and in 25\,\% of the evaluations, that the body is only in the sliding phase (Conclusion~2). When the large dataset was evaluated, in 70\,\% of the evaluations, participants came to Conclusion~3, and in 30\,\% to Conclusion~2. None of the participants concluded that the body is in a sticking phase in the corresponding time interval (Conclusion~1). A $\chi^2$-test of independence showed that the proportion of conclusions considered correct from a physics expert perspective (for the small dataset: Conclusion~2; for the large dataset: Conclusion~3) differed significantly between the two dataset conditions ($\chi^2(1) = 6.42$, $p = .011$).

\subsection{Data evaluation across participants: written justifications and visual attention \label{sec:Data evaluation across participants: written justifications and visual attention}}

Given that there is evidence that concurrent think-aloud can distort fixation data~\cite{Hahn2022,Becker2023}, we examined whether the fixation data collected without think-aloud differed from the fixation data collected using concurrent think-aloud. $\chi^2$ tests did not identify significant differences for the force data ($\chi^2(11) = 7.49,~p = 0.758$) or for the position data ($\chi^2(10) = 11.30,~p = 0.335$). However, given the limited sample size, the absence of significant effects should be interpreted with caution and does not necessarily indicate equivalence between conditions. Therefore, the data were analyzed together, while acknowledging that potential differences may have remained undetected.

\begin{figure*}
 \includegraphics[width=0.8\textwidth]{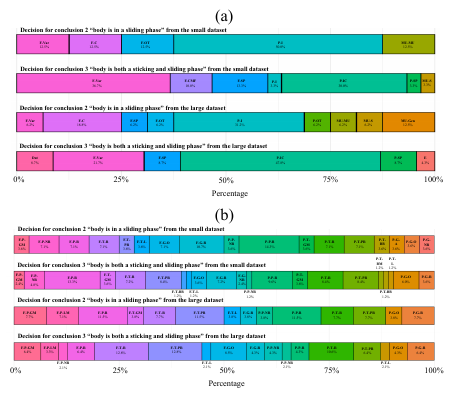}%
 \caption{Frequency of (a) elements referred to in the written explanations justifying the conclusion selected and (b) eye fixations areas in the eye-tracking fixation data that were taken into account in drawing a conclusion.\newline
Note. For the analysis of written justifications: Dat $=$ Measurement data in general, E $=$ Experimental setting, F-Var $=$ Variations in force, F-C $=$ Constant force, F-CMF $=$ Comparison of force data to maximum force measurement at 0.5\,s, F-SP $=$ Single measurements in force, F-OT $=$ Overall trend in force data, P-I $=$ Increasing position, P-IC $=$ Increasing and constant intervals in position data, P-SP $=$ Single measurements in position data, P-OT $=$ Overall trend in position data, MU-MU $=$ Measurement uncertainty of data, MU-S $=$ Measurement uncertainty resulting from sampling rate, MU-Gen $=$ Uncertainty in general. For the analysis of eye-tracking fixation data: The first element F (force) or P (position) refers to the measurand fixated. The second element P (point), T (trend), or G (general) refers to the point/interval that was fixed. The third element: GM $=$ global maximum, NR $=$ not in relevant time interval, R $=$ in relevant time interval, PR $=$ primary relevant time interval, L $=$ locally fixed time interval, BB $=$ beginning of movement, BM $=$ before movement, O $=$ overall. More details on the codes can be found in Appendix~\ref{sec:Coding manual for written justifications and think-aloud data} and Appendix~\ref{sec:Coding manual for eye-tracking fixation data}.
\label{fig:Frequency of (a) elements referred to in the written explanations justifying the conclusion selected and (b) eye fixations areas in the eye-tracking fixation data that were taken into account in drawing a conclusion.}}
 \end{figure*}

An overview of the distributions of coded categories in written justifications and eye-tracking data across all participants is shown in Figure~\ref{fig:Frequency of (a) elements referred to in the written explanations justifying the conclusion selected and (b) eye fixations areas in the eye-tracking fixation data that were taken into account in drawing a conclusion.}.

%\begin{widetext}
% \end{widetext}

Across all conditions, written justifications primarily refer to force-related and position-related aspects. For the small dataset, participants who chose Conclusion~2 more frequently refer to position-related aspects (50.0\,\%) than to force-related aspects (37.5\,\%), whereas participants who chose Conclusion~3 more frequently refer to force-related aspects (60.0\,\%) than to position-related aspects (36.6\,\%). For the large dataset, this finding is reversed: participants who chose Conclusion~2 refer more frequently to force-related aspects (37.4\,\%) than to position-related aspects (37.5\,\%), while participants who chose Conclusion~3 refer more frequently to position-related aspects (56.5\,\%) than to force-related aspects (30.4\,\%). References to measurement uncertainty in written justifications occurred in all conditions, with higher proportions for Conclusion~2 in the large dataset (24.9\,\%) and for Conclusion~2 in the small dataset (12.5\,\%) compared to the respective Conclusion~3 conditions (large dataset: 0\,\%, small dataset: 3.3\,\%).

Across all conditions, eye-tracking data show that fixation time is distributed across force-related and position-related areas, with a higher overall proportion of fixations on position data. For the small dataset, participants who chose Conclusion~2 show slightly higher proportions of fixations on position data (50.1\,\%) compared to force data (49.9\,\%), whereas participants who chose Conclusion~3 show a higher proportion of fixations on force data (55.3\,\%) than on position data (44.4\,\%). For the large dataset, fixation proportions on force data increased (56.9\,\% and 58.7\,\% for Conclusions 2 and 3), while fixations on position data decreased correspondingly (42.2\,\% and 40.5\,\%). Within the eye-tracking data, fixations occurred both in individual data points and on trends. For force-related fixations, the proportion of trend-based fixations increases from the small to the large dataset (e.g., F-T-: 14.3\,\% to 26.8\,\% for Conclusion~2), while fixations on global features decreases (e.g., F-G-: 17.8\,\% to 3.8\,\%). A comparable distribution is observed for position-related fixations, with trend-based fixations present across all conditions and slightly lower proportions in the large dataset.

\subsection{Data evaluation processes in the condition with ET-TA: integrating written justifications, visual attention, and think-aloud data \label{sec:Data evaluation processes in the subsample: integrating written justifications, visual attention, and think-aloud data}}

The distributions of coded categories in written justifications for the ET and the ET-TA conditions are shown in Figure~\ref{fig:Frequency of (a) elements referred to in the written explanations justifying the conclusion selected, (b) eye fixations areas in the eye-tracking fixation data that were taken into account in drawing a conclusion and, (c) identified references in the concurrent think-aloud data that were considered in drawing a conclusion.}. To determine whether the participants that underwent the condition data evaluation using think-aloud (ET-TA) differed from the overall sample, we conducted $\chi^2$-tests with simulated $p$-values based on 2000 replicates (due to small expected frequencies). The results did not reveal statistically significant differences in written justifications ($\chi^2 = 9.86,~p = 0.805$) or eye-fixations ($\chi^2 = 18.49,~p = 0.713$) between the conditions with and without think-aloud. However, given the small sample size and resulting limited statistical power, these findings should be interpreted cautiously. In addition, visual inspection of relative distributions served as a robustness check and confirmed a high degree of consistency across samples, with no qualitatively distinct findings emerging. Consequently, the condition with ET-TA replicates the findings observed in the full dataset reported in Section~\ref{sec:Differences between evaluating small and large datasets} for written justifications and eye-tracking data.

 \begin{figure*}
 %\includegraphics[width=0.55\textwidth]{Figure4.pdf}%
 %für doublespace
 \includegraphics[width=0.65\textwidth]{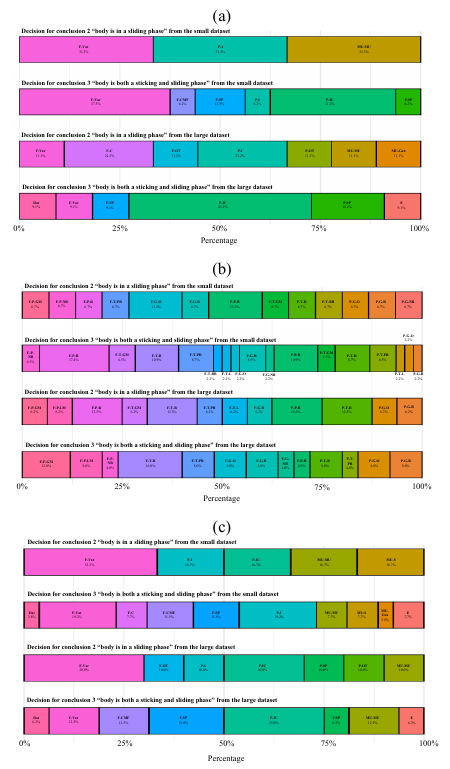}%
 \caption{Frequency of (a) elements referred to in the written explanations justifying the conclusion selected, (b) eye fixations areas in the eye-tracking fixation data that were taken into account in drawing a conclusion and, (c) identified references in the concurrent think-aloud data that were considered in drawing a conclusion.\newline
Note. For the analysis of written justifications: Dat $=$ Measurement data in general, E $=$ Experimental setting, F-Var $=$ Variations in force, F-C $=$ Constant force, F-CMF $=$ Comparison of force data to maximum force measurement at 0.5\,s, F-SP $=$ Single measurements in force, F-OT $=$ Overall trend in force data, P-I $=$ Increasing position, P-IC $=$ Increasing and constant intervals in position data, P-SP $=$ Single measurements in position data, P-OT $=$ Overall trend in position data, MU-MU $=$ Measurement uncertainty of data, MU-S $=$ Measurement uncertainty resulting from sampling rate, MU-Gen $=$ Uncertainty in general. For the analysis of eye-tracking fixation data: The first element F (force) or P (position) refers to the measurand fixated. The second element P (point), T (trend), or G (general) refers to the point/interval that was fixed. The third element: GM $=$ global maximum, NR $=$ not in relevant time interval, R $=$ in relevant time interval, PR $=$ primary relevant time interval, L $=$ locally fixed time interval, BB $=$ beginning of movement, BM $=$ before movement, O $=$ overall. More details on the codes can be found in Appendix~\ref{sec:Coding manual for written justifications and think-aloud data} and Appendix~\ref{sec:Coding manual for eye-tracking fixation data}.
\label{fig:Frequency of (a) elements referred to in the written explanations justifying the conclusion selected, (b) eye fixations areas in the eye-tracking fixation data that were taken into account in drawing a conclusion and, (c) identified references in the concurrent think-aloud data that were considered in drawing a conclusion.}}
 \end{figure*}

Think-aloud data show that participants primarily refer to force-related and position-related aspects across all conditions, which is in line with the distributions observed in the written justifications. For the small dataset, participants who chose Conclusion~2 refer equally to force-related aspects, position-related aspects, and measurement uncertainty (each 33.3\,\%), whereas participants who chose Conclusion~3 more frequently refer to force-related aspects (49.5\,\%) than to position-related aspects (19.2\,\%). For the large dataset, participants who chose Conclusion~2 refer more frequently to position-related aspects (50.0\,\%) than to force-related aspects (40.0\,\%), while participants chose for Conclusion~3 show a more balanced distribution across force-related (43.8\,\%) and position-related aspects (31.2\,\%). References to measurement uncertainty and experimental aspects are presented in the think-aloud data across all conditions, with varying proportions across conclusions and dataset sizes.

The integrated analysis of the three data sources indicates that participants employ different data evaluation strategies depending on the dataset they evaluated. For the large dataset (which also displayed more pronounced patterns), the data suggest a stronger focus on trend-based data evaluation, particularly for force-related data, as reflected in an increase of force-related trend fixations (18.6\,\% to 27.6\,\% in the eye-tracking data), while point-based fixations remain comparatively stable (20.9\,\% to 24.5\,\%). At the same time, position-related aspects become more prominent in both think-aloud data (26.3\,\% to 40.6\,\%) and written justifications (38.5\,\% to 48.5\,\%). In contrast, the evaluation of the small dataset is characterized by a stronger emphasis on force-related data (e.g., 44.8\,\% in written justifications) and a higher proportion of references to measurement uncertainty (e.g., 26.3\,\% in think-aloud data), indicating a more locally oriented and uncertainty-related evaluation strategy. Taken together, these findings suggest that participants in this study adapted their data evaluation strategies to the amount of data by shifting from more local and uncertainty-related processing toward more trend-based and integrative approaches.

Integrating written justifications, eye-tracking data, and think-aloud protocols further reveals systematic differences in participants’ data evaluation processes depending on dataset size. Across all process data, a selective integration of information can be identified: although eye-tracking fixations showed broad visual attention (e.g., on average 60.0\,\% of all fixations were force-related), only a subset of this information is verbalized (on average 41.7\,\%) and further reduced in written justifications (on average 38.0\,\%). This finding points to a multi-stage selection process from perception to verbalization to written argumentation. In line with the strategy differences described above, the relative emphasis on force- and position-related information shifted with dataset size, from a force-dominated evaluation in the small dataset (e.g., 56.2\,\% vs. 18.2\,\% in written justifications for Conclusion~3) to a stronger emphasis on position in the large dataset (e.g., 43.6\,\% vs. 63.7\,\%). Finally, references to measurement uncertainty were more prominent during the evaluation of the small dataset (e.g., 33.4\,\% in think-aloud data for Conclusion~2) and less frequent for the large dataset, indicating differences in how uncertainty is processed during data evaluation. Taken together, these findings indicate that within this sample, participants evaluated datasets of different sizes by adapting their strategies toward more trend-based, integrative processing for larger datasets, while smaller datasets are associated with more local and uncertainty-related processing.

\section{Discussion \label{sec:Discussion}}

\subsection{Differences between evaluating small and large datasets \label{sec:Differences between evaluating small and large datasets}}

The aim of this study was to investigate how students evaluate different amounts of measurement data when drawing conclusions about physical phenomena. The findings indicate systematic differences within the present sample in how students evaluate small and large datasets, both in terms of the conclusions drawn and the underlying data evaluation processes. The observed differences between conditions can be interpreted in light of the fact that dataset size and pattern visibility were manipulated simultaneously.

First, evaluating the large dataset (which also made underlying patterns more visible) was associated with an increase of the likelihood that participants selected the conclusion consistent with a physics expert interpretation. A significantly higher proportion of participants selected the expert-supported conclusion when evaluating the large dataset compared to the small dataset. In contrast, the evaluation of the small dataset led to a more distributed selection of conclusions, with participants more frequently selecting conclusions that were not consistent with a physics expert interpretation. This finding is noteworthy, as theoretical accounts often associate larger datasets with increased complexity due to heterogeneity and the need for data evaluation~\cite{Schultheis2015,Benz2025,Fielding2025,Kastens2015,Kjelvik2019,Rosenberg2022,Wilkerson2025,Lee2018}. However, the present results suggest that, in the present context, the large dataset can also help clarify ambiguities, particularly where it makes patterns in the data more visible and clearer.

Second, the think-aloud data from the participants' (and, to some extent, participants' written justifications) show that, when evaluating the small dataset, participants were often less confident due to the limited number of data points, as reflected in frequent references to aspects of uncertainty, and they articulated a ``need for data." At the same time, they tended to focus on fluctuations in individual data points rather than on overarching patterns, which predominantly led to overinterpretation or overweighting of individual data points and thus contributed to the observed ambiguity of the conclusions. Taken together, these results point to a possible evaluation strategy employed by the participants when dealing with few data points: Although the participants apparently recognized that the available data was limited and articulated a need for more data, they did not necessarily respond with more cautious interpretations. Instead, they compensated for the lack of data by assigning greater interpretive weight to individual observations than would have been justified given the dataset. This interpretation is consistent with prior work showing that small datasets can promote a focus on individual observations and limit opportunities for pattern recognition~\cite{Kastens2015,Kjelvik2019,Rosenberg2022,Lee2018,Benz2025}, while also increasing the salience of uncertainty~\cite{Masnick2008}. Furthermore, it is plausible that this strategy contributes to the distribution of supported conclusions observed in previous research with high school students~\cite{Benz2025,Benz2024}. However, this interpretation is not directly supported by the data collected in the present study and should therefore be understood as a hypothesis for future research rather than as a confirmed explanatory mechanism. 

Third, evaluating large datasets is associated with a stronger emphasis on pattern- or trend-based evaluation. Across all types of process data collected, participants more frequently attended to and referred to trends and intervals when dealing with the large dataset, particularly for force-related data. This shift suggests that, in the present context, the large dataset supported the integration of multiple data points into coherent patterns, thereby supporting more global interpretations of the represented physical phenomenon. This finding aligns with theoretical accounts emphasizing that large datasets can reveal underlying structures through aggregation and integrative analysis~\cite{Kastens2015,Resnick2018,Wilkerson2025,Benz2024}. Furthermore, it appears plausible that students may apply cognitive summarizing mechanisms~\cite{Masnick2022} more implicitly when dealing with larger datasets than with smaller ones. Particularly in tasks where students must deal with measurement uncertainties~\cite{Morris2022,Kok2024}, the potential of large datasets may become apparent, as students increasingly rely on such summarizing mechanisms to evaluate data and place less emphasis on individual measurement values.

Fourth, evaluating small datasets is characterized by more local and uncertainty-related data evaluations. Participants more frequently focused on individual data points and referred to measurement uncertainty when working with the small dataset. This suggests that, in the present context, the small dataset provided fewer opportunities to identify patterns and increased the salience of uncertainty, leading to more cautious and locally oriented evaluation strategies. This observation is consistent with theoretical perspectives that highlight how limited data can amplify uncertainty and constrain broader interpretation~\cite{Kastens2015,Kjelvik2019,Rosenberg2022,Lee2018,Masnick2008}. Possible underlying causes (e.g., ``need for data", overweighting/overinterpreting of single data points) are discussed in the two paragraphs above.

Finally, the findings indicate that, when evaluating the large dataset, participants more frequently attended to and referred to trends and intervals rather than individual data points. In particular, the increase in trend-based fixations and references to broader patterns suggests a shift toward more integrative forms of data evaluation. One possible interpretation of this finding is that participants organize and reduce the available information by relating multiple data points to one another. In this sense, the observed evaluation strategy may reflect what could be described as summarizing or~``chunking" mechanisms, in which information is aggregated across data points to form more coherent representations of the underlying phenomenon~\cite{Masnick2022}. However, these mechanisms were not directly measured in the present study and should therefore be interpreted as a plausible explanatory account rather than as an empirically established process. This interpretation is backed by data from prior research from the same physical context in which high-school students evaluated different-sized datasets presented in diagrams and reported constant cognitive load, leading to the assumption that students aggregated multiple data points into features (e.g., trends, patterns)~\cite{Benz2024}. In addition, this interpretation is consistent with research from the field suggesting that larger datasets can promote integrative evaluation and the use of cognitive strategies aimed at structuring and reducing information~\cite{Kastens2015,Resnick2018,Wilkerson2025,Benz2024,Schultheis2015,Benz2025,Fielding2025,Kjelvik2019,Rosenberg2022,Lee2018}. More generally, these interpretations highlight the need for future studies that directly assess underlying cognitive mechanisms during data evaluation; for instance, by retrospective interviews.

Overall, these findings indicate that differences in dataset size, particularly when accompanied by differences in pattern visibility, are associated with systematic differences in how students evaluate and interpret data. In the present study, increasing dataset size was accompanied by changes in the visibility of underlying patterns, making it difficult to disentangle the effects of data quantity from those of pattern structure. Accordingly, the observed shift from more local and uncertainty-related data evaluation toward more pattern-based and integrative approaches should be understood as context-dependent rather than as a general effect of increasing data quantity. From a physics education perspective, this suggests that larger datasets may hold pedagogical potential particularly in contexts where they make relevant patterns more visible and support integrative forms of data evaluation.

\subsection{Data evaluation processes across different amounts of measurement data \label{sec:Data evaluation processes across different amounts of measurement data}}

Beyond the differences between small and large datasets, the findings indicate that data evaluation is not a direct process of extracting meaning from measurement data, but rather a selective, multi-stage process. Across all conditions, integrating eye-tracking data, think-aloud protocols, and written justifications showed a systematic reduction of information from visual attention to verbalization to written justifications. While participants visually attended to a broad range of elements in the diagrams, only a subset of this information was verbalized during think-aloud, and an even smaller subset was incorporated into written justifications. This finding suggests that data evaluation involves successive stages of selection and transformation, in which initially available information is filtered and restructured before being used to justify conclusions. In line with prior work showing that students’ came to incorrect conclusions from correct reasons and vice versa~\cite{Kok2024,Buffler2001,Sere2001}, these findings highlight that outcome measures alone are insufficient and underscore the value of integrating multiple process measures to capture students’ data-based reasoning more comprehensively. In turn, this is consistent with prior work showing that visually similar attention patterns can be associated with different underlying reasoning processes, and that combining eye-tracking data with think-aloud protocols provides a more comprehensive account of data evaluation~\cite{Klassen2026}. Although no significant differences were found between eye-tracking conditions with and without concurrent think-aloud, this result should be interpreted with caution. Given the relatively small sample size, the study may not have been sufficiently powered to detect subtle differences between conditions. Consequently, potential influences of the think-aloud procedure on visual attention and reasoning processes cannot be fully ruled out.

The findings further emphasize the central role of visual attention in data evaluation, while also illustrating its limitations. Consistent with prior research and the \textit{eye-mind hypothesis}~\cite{Carpenter1989}, fixation findings provide insight into which elements of a diagram are cognitively processed during data evaluation. In line with prior research~\cite{Becker2023,Hahn2022,Madsen2012,Madsen2013}, participants attended to both relevant (measurement data points/trends/patterns within the time interval of interest) and non-relevant or less irrelevant areas (measurement data points/trends/patterns outside of the time interval of interest) of the diagrams, indicating that visual attention is closely linked to the data evaluation process. However, the observed discrepancies between fixation data, think-aloud data, and written justifications show that visual attention alone does not fully account for how information is interpreted and used for reasoning. Participants sometimes fixated relevant areas without incorporating them into their evaluation processes (not identified in think-aloud data) or focused on aspects that were not reflected in their written justifications. This finding supports concerns raised in the literature regarding the limits of interpreting eye-tracking data in isolation~\cite{Kok2016} and highlights the need to complement visual attention measures with additional data sources to better capture underlying cognitive processes.

Finally, the findings suggest that data evaluation is shaped by perceived uncertainty and the strategies students employ to manage it. References to measurement uncertainty in both think-aloud data and written justifications indicate that participants actively consider the reliability and interpretability of measurement data during data evaluation. This observation aligns with theoretical accounts emphasizing the role of epistemic uncertainty in working with data, particularly in less idealized settings~\cite{Kastens2015,Lee2018,Fielding2025,Wilkerson2025}. At the same time, the findings show that participants do not evaluate all available information equally, but instead selectively focus on specific aspects, such as trends or individual data points, depending on the situation. This selective processing reflects strategy use, which has been linked to heuristic reasoning and the handling of complex or uncertain data in prior work~\cite{Baur2018,GarciaMila2017,Ludwig2021,Lenz2025}. The results also tentatively suggest that perceived uncertainty is linked to the use of specific interpretative strategies; in particular, uncertainty may be accompanied by a tendency to compensate for limited data by assigning greater weight to individual data points. Further research is needed to identify such relations. Taken together, these findings across conditions suggest that data evaluation is not only a perceptual process, but also a strategy-dependent activity in which students actively manage uncertainty by selecting, weighting, and integrating information.

\subsection{Limitations \label{sec:Limitations}}

In addition to the limitations that stem from the relatively small sample of university students this study is based upon and that have been discusses previously, there are two addition limitations worth noting: First, the study focused on a single experimental context, namely a friction experiment. This context is highly specific in that increasing the sampling rate not only changed the amount of data but also the pattern structure of the datasets. As a result, the comparison between conditions does not isolate dataset size alone but also includes differences in the data patterns that became visible. Future research should therefore examine a broader range of contexts and ways of varying data quantity.

Second, the sequential evaluation of two datasets represents a substantial limitation, as participants may have developed an understanding of the underlying physical phenomenon during the first task, which could have influenced their interpretation of the second dataset. Although presentation order was counterbalanced, working with one dataset may have shaped how participants approached the second one, for example through strategy transfer or confirmation tendencies. Such carry-over effects cannot be disentangled from the effects attributed to dataset size and may have contributed to the observed differences between conditions. Future studies could address this issue using between-subject designs or tasks that require direct comparison of datasets.

\section{Conclusions \label{sec:Conclusions}}

This study examined how students evaluate different amounts of measurement data when drawing conclusions about physical phenomena. Although the study should be understood as exploratory and interpreted in light of its limitations, the findings suggest that dataset size, in interaction with the visibility of patterns in the data, shapes students' data evaluation processes in the present context, rather than merely changing the amount of available information. Evaluating larger datasets led to more unambiguous conclusions and was associated with more trend-based and integrative reasoning. Evaluating smaller datasets, in contrast, led to more distributed conclusions, a stronger focus on individual data points, and more references to measurement uncertainty. These results suggest that students adapt their data evaluation processes to dataset size by shifting between more local and more integrative forms of reasoning. At the same time, the findings indicate that evaluating larger datasets requires specific data-handling strategies, such as structuring and summarizing measurement data to identify relevant patterns. Without such strategies, increasing the amount of data does not necessarily lead to more appropriate conclusions. Overall, the study highlights that effective data evaluation depends not only on how much data are available, but on how students interpret, structure, and integrate measurement data during reasoning. As an exploratory study, this work revealed several promising directions for further research on students' data evaluation processes.

\begin{acknowledgments}
We thank Katja Scholz for her support in double-coding the data.
\end{acknowledgments}

\bibliography{PRPER-Lit}

\begin{thebibliography}{64}%
\makeatletter
\providecommand \@ifxundefined [1]{%
 \@ifx{#1\undefined}
}%
\providecommand \@ifnum [1]{%
 \ifnum #1\expandafter \@firstoftwo
 \else \expandafter \@secondoftwo
 \fi
}%
\providecommand \@ifx [1]{%
 \ifx #1\expandafter \@firstoftwo
 \else \expandafter \@secondoftwo
 \fi
}%
\providecommand \natexlab [1]{#1}%
\providecommand \enquote  [1]{``#1''}%
\providecommand \bibnamefont  [1]{#1}%
\providecommand \bibfnamefont [1]{#1}%
\providecommand \citenamefont [1]{#1}%
\providecommand \href@noop [0]{\@secondoftwo}%
\providecommand \href [0]{\begingroup \@sanitize@url \@href}%
\providecommand \@href[1]{\@@startlink{#1}\@@href}%
\providecommand \@@href[1]{\endgroup#1\@@endlink}%
\providecommand \@sanitize@url [0]{\catcode `\\12\catcode `\$12\catcode `\&12\catcode `\#12\catcode `\^12\catcode `\_12\catcode `\%12\relax}%
\providecommand \@@startlink[1]{}%
\providecommand \@@endlink[0]{}%
\providecommand \url  [0]{\begingroup\@sanitize@url \@url }%
\providecommand \@url [1]{\endgroup\@href {#1}{\urlprefix }}%
\providecommand \urlprefix  [0]{URL }%
\providecommand \Eprint [0]{\href }%
\providecommand \doibase [0]{https://doi.org/}%
\providecommand \selectlanguage [0]{\@gobble}%
\providecommand \bibinfo  [0]{\@secondoftwo}%
\providecommand \bibfield  [0]{\@secondoftwo}%
\providecommand \translation [1]{[#1]}%
\providecommand \BibitemOpen [0]{}%
\providecommand \bibitemStop [0]{}%
\providecommand \bibitemNoStop [0]{.\EOS\space}%
\providecommand \EOS [0]{\spacefactor3000\relax}%
\providecommand \BibitemShut  [1]{\csname bibitem#1\endcsname}%
\let\auto@bib@innerbib\@empty
%</preamble>
\bibitem [{\citenamefont {{National Academies of Sciences, Engineering and Medicine}}(2019)}]{NationalAcademies}%
  \BibitemOpen
  \bibfield  {author} {\bibinfo {author} {\bibnamefont {{National Academies of Sciences, Engineering and Medicine}}},\ }\href@noop {} {\emph {\bibinfo {title} {Science and {Engineering} for {Grades} 6-10}}}\ (\bibinfo  {publisher} {The National Academies Press},\ \bibinfo {year} {2019})\BibitemShut {NoStop}%
\bibitem [{\citenamefont {Thornton}(1992)}]{Thornton1992}%
  \BibitemOpen
  \bibfield  {author} {\bibinfo {author} {\bibfnamefont {R.~K.}\ \bibnamefont {Thornton}},\ }\bibfield  {title} {\bibinfo {title} {Tools for {Scientific} {Thinking}: {Learning} {Physical} {Concepts} with {Real}-{Time} {Laboratory} {Measurement} {Tools}},\ }in\ \href {https://doi.org/10.1007/978-3-642-77750-9_12} {\emph {\bibinfo {booktitle} {New {Directions} in {Educational} {Technology}}}},\ \bibinfo {editor} {edited by\ \bibinfo {editor} {\bibfnamefont {E.}~\bibnamefont {Scanlon}}\ and\ \bibinfo {editor} {\bibfnamefont {T.}~\bibnamefont {O'Shea}}}\ (\bibinfo  {publisher} {Springer},\ \bibinfo {address} {Berlin, Heidelberg},\ \bibinfo {year} {1992})\ pp.\ \bibinfo {pages} {139--151}\BibitemShut {NoStop}%
\bibitem [{\citenamefont {Volkwyn}(2005)}]{Volkwyn2005}%
  \BibitemOpen
  \bibfield  {author} {\bibinfo {author} {\bibfnamefont {T.~S.}\ \bibnamefont {Volkwyn}},\ }\emph {\bibinfo {title} {First year students' understanding of measurement in physics laboratory work}},\ \href@noop {} {\bibinfo {type} {Dissertation}},\ \bibinfo  {school} {University of Cape Town}, \bibinfo {address} {Cape Town} (\bibinfo {year} {2005})\BibitemShut {NoStop}%
\bibitem [{\citenamefont {{National Research Council}}(2012)}]{NRC2012}%
  \BibitemOpen
  \bibfield  {author} {\bibinfo {author} {\bibnamefont {{National Research Council}}},\ }\href {https://doi.org/10.17226/13165} {\emph {\bibinfo {title} {A {Framework} for {K}-12 {Science} {Education}: {Practices}, {Crosscutting} {Concepts}, and {Core} {Ideas}}}}\ (\bibinfo  {publisher} {The National Academies Press},\ \bibinfo {address} {Washington, DC},\ \bibinfo {year} {2012})\BibitemShut {NoStop}%
\bibitem [{\citenamefont {{NGSS Lead States}}(2013)}]{NGSS2013}%
  \BibitemOpen
  \bibfield  {author} {\bibinfo {author} {\bibnamefont {{NGSS Lead States}}},\ }\href@noop {} {\emph {\bibinfo {title} {Next {Generation} {Science} {Standards}: {For} {States}, {By} {States}}}},\ \bibinfo {type} {Tech. Rep.}\ (\bibinfo  {institution} {The National Academies Press},\ \bibinfo {address} {Washington, DC},\ \bibinfo {year} {2013})\BibitemShut {NoStop}%
\bibitem [{\citenamefont {KMK}(2024)}]{KMK2024}%
  \BibitemOpen
  \bibfield  {author} {\bibinfo {author} {\bibnamefont {KMK}},\ }\href@noop {} {\emph {\bibinfo {title} {Bildungsstandards im {Fach} {Physik} für die {Allgemeine} {Hochschulreife} ({Beschluss} der {Kultusministerkonferenz} vom 18.06.2020) [{Educational} standards in physics for the {Allgemeine} {Hochschulreife} (resolution of the {Standing} {Conference} of the {Ministers} of {Education} and {Cultural} {Affairs} of the {Länder} in the {Federal} {Republic} of {Germany} of 18.06.2020)]}}}\ (\bibinfo  {publisher} {Carl Link},\ \bibinfo {year} {2024})\BibitemShut {NoStop}%
\bibitem [{\citenamefont {{Department of Education}}(2014)}]{UK2014}%
  \BibitemOpen
  \bibfield  {author} {\bibinfo {author} {\bibnamefont {{Department of Education}}},\ }\href@noop {} {\emph {\bibinfo {title} {Science programmes of study: key stage 4}}},\ \bibinfo {type} {Tech. Rep.}\ \bibinfo {number} {DFE-00677-2014}\ (\bibinfo {year} {2014})\BibitemShut {NoStop}%
\bibitem [{\citenamefont {Boyd}\ and\ \citenamefont {Crawford}(2012)}]{Boyd2012}%
  \BibitemOpen
  \bibfield  {author} {\bibinfo {author} {\bibfnamefont {D.}~\bibnamefont {Boyd}}\ and\ \bibinfo {author} {\bibfnamefont {K.}~\bibnamefont {Crawford}},\ }\bibfield  {title} {\bibinfo {title} {Critical {Questions} for {Big} {Data}. {Provocations} for a cultural, technological, and scholarly phenomenon},\ }\href {https://doi.org/10.1080/1369118X.2012.678878} {\bibfield  {journal} {\bibinfo  {journal} {Information, Communication \& Society}\ }\textbf {\bibinfo {volume} {15}},\ \bibinfo {pages} {662} (\bibinfo {year} {2012})}\BibitemShut {NoStop}%
\bibitem [{\citenamefont {Duggan}\ and\ \citenamefont {Gott}(2002)}]{Duggan2002}%
  \BibitemOpen
  \bibfield  {author} {\bibinfo {author} {\bibfnamefont {S.}~\bibnamefont {Duggan}}\ and\ \bibinfo {author} {\bibfnamefont {R.}~\bibnamefont {Gott}},\ }\bibfield  {title} {\bibinfo {title} {What sort of science education do we really need?},\ }\href {https://doi.org/10.1080/09500690110110133} {\bibfield  {journal} {\bibinfo  {journal} {International Journal of Science Education}\ }\textbf {\bibinfo {volume} {24}},\ \bibinfo {pages} {661} (\bibinfo {year} {2002})}\BibitemShut {NoStop}%
\bibitem [{\citenamefont {Kjelvik}\ and\ \citenamefont {Schultheis}(2019)}]{Kjelvik2019}%
  \BibitemOpen
  \bibfield  {author} {\bibinfo {author} {\bibfnamefont {M.~K.}\ \bibnamefont {Kjelvik}}\ and\ \bibinfo {author} {\bibfnamefont {E.~H.}\ \bibnamefont {Schultheis}},\ }\bibfield  {title} {\bibinfo {title} {Getting {Messy} with {Authentic} {Data}: {Exploring} the {Potential} of {Using} {Data} from {Scientific} {Research} to {Support} {Student} {Data} {Literacy}},\ }\href {https://doi.org/10.1187/cbe.18-02-0023} {\bibfield  {journal} {\bibinfo  {journal} {CBE - Life Sciences Education}\ }\textbf {\bibinfo {volume} {18}},\ \bibinfo {pages} {es2} (\bibinfo {year} {2019})}\BibitemShut {NoStop}%
\bibitem [{\citenamefont {Rosenberg}\ \emph {et~al.}(2022)\citenamefont {Rosenberg}, \citenamefont {Schultheis}, \citenamefont {Kjelvik}, \citenamefont {Reedy},\ and\ \citenamefont {Sultana}}]{Rosenberg2022}%
  \BibitemOpen
  \bibfield  {author} {\bibinfo {author} {\bibfnamefont {J.~M.}\ \bibnamefont {Rosenberg}}, \bibinfo {author} {\bibfnamefont {E.~H.}\ \bibnamefont {Schultheis}}, \bibinfo {author} {\bibfnamefont {M.~K.}\ \bibnamefont {Kjelvik}}, \bibinfo {author} {\bibfnamefont {A.}~\bibnamefont {Reedy}},\ and\ \bibinfo {author} {\bibfnamefont {O.}~\bibnamefont {Sultana}},\ }\bibfield  {title} {\bibinfo {title} {Big data, big changes? {The} technologies and sources of data used in science classrooms},\ }\href {https://doi.org/10.1111/bjet.13245} {\bibfield  {journal} {\bibinfo  {journal} {British Journal of Educational Technology}\ }\textbf {\bibinfo {volume} {53}},\ \bibinfo {pages} {1179} (\bibinfo {year} {2022})}\BibitemShut {NoStop}%
\bibitem [{\citenamefont {Benz}\ and\ \citenamefont {El~Hamdani-Ludwig}(2026)}]{Benz2026a}%
  \BibitemOpen
  \bibfield  {author} {\bibinfo {author} {\bibfnamefont {G.}~\bibnamefont {Benz}}\ and\ \bibinfo {author} {\bibfnamefont {T.}~\bibnamefont {El~Hamdani-Ludwig}},\ }\bibfield  {title} {\bibinfo {title} {Promoting (pre-service) physics teachers’ competencies in the field of digital measurement in laboratory work},\ }\href {https://doi.org/10.1088/1361-6404/ae4798} {\bibfield  {journal} {\bibinfo  {journal} {European Journal of Physics}\ }\textbf {\bibinfo {volume} {47}},\ \bibinfo {pages} {025708} (\bibinfo {year} {2026})}\BibitemShut {NoStop}%
\bibitem [{\citenamefont {Benz}(2024)}]{Benz2024}%
  \BibitemOpen
  \bibfield  {author} {\bibinfo {author} {\bibfnamefont {G.}~\bibnamefont {Benz}},\ }\emph {\bibinfo {title} {``{Big} {Data}" in {Inquiry}-{Based} {Learning} of {Science}}},\ \href {https://phka.bsz-bw.de/frontdoor/index/index/docId/659} {\bibinfo {type} {Dissertation}},\ \bibinfo  {school} {Karlsruhe University of Education}, \bibinfo {address} {Karlsruhe} (\bibinfo {year} {2024})\BibitemShut {NoStop}%
\bibitem [{\citenamefont {Benz}\ \emph {et~al.}(2025)\citenamefont {Benz}, \citenamefont {Ludwig},\ and\ \citenamefont {Vorholzer}}]{Benz2025}%
  \BibitemOpen
  \bibfield  {author} {\bibinfo {author} {\bibfnamefont {G.}~\bibnamefont {Benz}}, \bibinfo {author} {\bibfnamefont {T.}~\bibnamefont {Ludwig}},\ and\ \bibinfo {author} {\bibfnamefont {A.}~\bibnamefont {Vorholzer}},\ }\bibfield  {title} {\bibinfo {title} {Does {Size} {Matter}? {Impact} of {Handling} {Diagrams} {Presenting} {Different} {Amounts} of {Data} on {Students}' {Arguments} in {Educational} {Lab} {Settings}},\ }\href {https://doi.org/10.1002/sce.21985} {\bibfield  {journal} {\bibinfo  {journal} {Science Education}\ }\textbf {\bibinfo {volume} {109}},\ \bibinfo {pages} {1669} (\bibinfo {year} {2025})}\BibitemShut {NoStop}%
\bibitem [{\citenamefont {Benz}\ \emph {et~al.}(2022)\citenamefont {Benz}, \citenamefont {Buhlinger},\ and\ \citenamefont {Ludwig}}]{Benz2022}%
  \BibitemOpen
  \bibfield  {author} {\bibinfo {author} {\bibfnamefont {G.}~\bibnamefont {Benz}}, \bibinfo {author} {\bibfnamefont {C.}~\bibnamefont {Buhlinger}},\ and\ \bibinfo {author} {\bibfnamefont {T.}~\bibnamefont {Ludwig}},\ }\bibfield  {title} {\bibinfo {title} {`{Big} data' in physics education: discovering the stick-slip effect through a high sample rate},\ }\href {https://doi.org/10.1088/1361-6552/ac59cb} {\bibfield  {journal} {\bibinfo  {journal} {Physics Education}\ }\textbf {\bibinfo {volume} {57}},\ \bibinfo {pages} {045004} (\bibinfo {year} {2022})}\BibitemShut {NoStop}%
\bibitem [{\citenamefont {Kastens}\ \emph {et~al.}(2015)\citenamefont {Kastens}, \citenamefont {Krumhansl},\ and\ \citenamefont {Baker}}]{Kastens2015}%
  \BibitemOpen
  \bibfield  {author} {\bibinfo {author} {\bibfnamefont {K.}~\bibnamefont {Kastens}}, \bibinfo {author} {\bibfnamefont {R.}~\bibnamefont {Krumhansl}},\ and\ \bibinfo {author} {\bibfnamefont {I.}~\bibnamefont {Baker}},\ }\bibfield  {title} {\bibinfo {title} {Thinking big: {Transitioning} your students from working with small, student-collected data sets toward ``big data"},\ }\href {www.jstor.com/stable/43683256} {\bibfield  {journal} {\bibinfo  {journal} {The Science Teacher}\ }\textbf {\bibinfo {volume} {82}},\ \bibinfo {pages} {25} (\bibinfo {year} {2015})}\BibitemShut {NoStop}%
\bibitem [{\citenamefont {Schultheis}\ and\ \citenamefont {Kjelvik}(2015)}]{Schultheis2015}%
  \BibitemOpen
  \bibfield  {author} {\bibinfo {author} {\bibfnamefont {E.~H.}\ \bibnamefont {Schultheis}}\ and\ \bibinfo {author} {\bibfnamefont {M.~K.}\ \bibnamefont {Kjelvik}},\ }\bibfield  {title} {\bibinfo {title} {Data {Nuggets}: {Bringing} {Real} {Data} into the {Classroom} to {Unearth} {Students}' {Quantitative} \& {Inquiry} {Skills}},\ }\href {https://doi.org/10.1525/abt.2015.77.1.4} {\bibfield  {journal} {\bibinfo  {journal} {American Biology Teacher}\ }\textbf {\bibinfo {volume} {77}},\ \bibinfo {pages} {19} (\bibinfo {year} {2015})}\BibitemShut {NoStop}%
\bibitem [{\citenamefont {Fielding}\ \emph {et~al.}(2025)\citenamefont {Fielding}, \citenamefont {Makar},\ and\ \citenamefont {Ben-Zvi}}]{Fielding2025}%
  \BibitemOpen
  \bibfield  {author} {\bibinfo {author} {\bibfnamefont {J.}~\bibnamefont {Fielding}}, \bibinfo {author} {\bibfnamefont {K.}~\bibnamefont {Makar}},\ and\ \bibinfo {author} {\bibfnamefont {D.}~\bibnamefont {Ben-Zvi}},\ }\bibfield  {title} {\bibinfo {title} {Developing students' reasoning with data and data-ing},\ }\href {https://doi.org/10.1007/s11858-025-01671-6} {\bibfield  {journal} {\bibinfo  {journal} {ZDM - Mathematics Education}\ }\textbf {\bibinfo {volume} {57}},\ \bibinfo {pages} {1} (\bibinfo {year} {2025})}\BibitemShut {NoStop}%
\bibitem [{\citenamefont {Wilkerson}\ \emph {et~al.}(2025)\citenamefont {Wilkerson}, \citenamefont {Erickson}, \citenamefont {Lee},\ and\ \citenamefont {Finzer}}]{Wilkerson2025}%
  \BibitemOpen
  \bibfield  {author} {\bibinfo {author} {\bibfnamefont {M.~H.}\ \bibnamefont {Wilkerson}}, \bibinfo {author} {\bibfnamefont {T.}~\bibnamefont {Erickson}}, \bibinfo {author} {\bibfnamefont {H.}~\bibnamefont {Lee}},\ and\ \bibinfo {author} {\bibfnamefont {W.}~\bibnamefont {Finzer}},\ }\bibfield  {title} {\bibinfo {title} {How to be ``{Choosy}": {Wrangling} big datasets for the classroom},\ }\href {https://doi.org/10.1111/test.70022} {\bibfield  {journal} {\bibinfo  {journal} {Teaching Statistics}\ ,\ \bibinfo {pages} {1}} (\bibinfo {year} {2025})}\BibitemShut {NoStop}%
\bibitem [{\citenamefont {Lee}\ and\ \citenamefont {Wilkerson}(2018)}]{Lee2018}%
  \BibitemOpen
  \bibfield  {author} {\bibinfo {author} {\bibfnamefont {V.~R.}\ \bibnamefont {Lee}}\ and\ \bibinfo {author} {\bibfnamefont {M.~H.}\ \bibnamefont {Wilkerson}},\ }\bibfield  {title} {\bibinfo {title} {Data {Use} by {Middle} and {Secondary} {Students} in the {Digital} {Age}: {A} {Status} {Report} and {Future} {Prospects}},\ }\href@noop {} {\bibfield  {journal} {\bibinfo  {journal} {Commissioned Paper for the National Academies of Sciences, Engineering, and Medicine, Board on Science Education, Committee on Science Investigations and Engineering Design for Grades 6-12}\ } (\bibinfo {year} {2018})}\BibitemShut {NoStop}%
\bibitem [{\citenamefont {Resnick}\ \emph {et~al.}(2018)\citenamefont {Resnick}, \citenamefont {Kastens},\ and\ \citenamefont {Shipley}}]{Resnick2018}%
  \BibitemOpen
  \bibfield  {author} {\bibinfo {author} {\bibfnamefont {I.}~\bibnamefont {Resnick}}, \bibinfo {author} {\bibfnamefont {K.~A.}\ \bibnamefont {Kastens}},\ and\ \bibinfo {author} {\bibfnamefont {T.~F.}\ \bibnamefont {Shipley}},\ }\bibfield  {title} {\bibinfo {title} {How students reason about visualizations from large professionally collected data sets: {A} study of students approaching the threshold of data proficiency},\ }\href {https://doi.org/10.1080/10899995.2018.1411724} {\bibfield  {journal} {\bibinfo  {journal} {Journal of Geoscience Education}\ }\textbf {\bibinfo {volume} {66}},\ \bibinfo {pages} {55} (\bibinfo {year} {2018})}\BibitemShut {NoStop}%
\bibitem [{\citenamefont {Bowen}\ and\ \citenamefont {Roth}(2007)}]{Bowen2007}%
  \BibitemOpen
  \bibfield  {author} {\bibinfo {author} {\bibfnamefont {G.~M.}\ \bibnamefont {Bowen}}\ and\ \bibinfo {author} {\bibfnamefont {W.-M.}\ \bibnamefont {Roth}},\ }\bibfield  {title} {\bibinfo {title} {The {Practice} of {Field} {Ecology}: {Insights} for {Science} {Education}},\ }\href {https://doi.org/10.1007/s11165-006-9021-x} {\bibfield  {journal} {\bibinfo  {journal} {Research in Science Education}\ }\textbf {\bibinfo {volume} {37}},\ \bibinfo {pages} {171} (\bibinfo {year} {2007})}\BibitemShut {NoStop}%
\bibitem [{\citenamefont {Schang}\ \emph {et~al.}(2023)\citenamefont {Schang}, \citenamefont {Dew}, \citenamefont {Stump}, \citenamefont {Holmes},\ and\ \citenamefont {Passante}}]{Schang2023}%
  \BibitemOpen
  \bibfield  {author} {\bibinfo {author} {\bibfnamefont {A.}~\bibnamefont {Schang}}, \bibinfo {author} {\bibfnamefont {M.}~\bibnamefont {Dew}}, \bibinfo {author} {\bibfnamefont {E.~M.}\ \bibnamefont {Stump}}, \bibinfo {author} {\bibfnamefont {N.~G.}\ \bibnamefont {Holmes}},\ and\ \bibinfo {author} {\bibfnamefont {G.}~\bibnamefont {Passante}},\ }\bibfield  {title} {\bibinfo {title} {New perspectives on student reasoning about measurement uncertainty: {More} or better data},\ }\href {https://doi.org/10.1103/PhysRevPhysEducRes.19.020105} {\bibfield  {journal} {\bibinfo  {journal} {Physical Review Physics Education Research}\ }\textbf {\bibinfo {volume} {19}},\ \bibinfo {pages} {020105} (\bibinfo {year} {2023})}\BibitemShut {NoStop}%
\bibitem [{\citenamefont {Masnick}\ and\ \citenamefont {Morris}(2008)}]{Masnick2008}%
  \BibitemOpen
  \bibfield  {author} {\bibinfo {author} {\bibfnamefont {A.~M.}\ \bibnamefont {Masnick}}\ and\ \bibinfo {author} {\bibfnamefont {B.~J.}\ \bibnamefont {Morris}},\ }\bibfield  {title} {\bibinfo {title} {Investigating the {Development} of {Data} {Evaluation}: {The} {Role} of {Data} {Characteristics}},\ }\href {https://doi.org/10.1111/j.1467-8624.2008.01174.x} {\bibfield  {journal} {\bibinfo  {journal} {Child Development}\ }\textbf {\bibinfo {volume} {79}},\ \bibinfo {pages} {1032} (\bibinfo {year} {2008})}\BibitemShut {NoStop}%
\bibitem [{\citenamefont {Boaventura}\ \emph {et~al.}(2013)\citenamefont {Boaventura}, \citenamefont {Faria}, \citenamefont {Chagas},\ and\ \citenamefont {Galvāo}}]{Boaventura2013}%
  \BibitemOpen
  \bibfield  {author} {\bibinfo {author} {\bibfnamefont {D.}~\bibnamefont {Boaventura}}, \bibinfo {author} {\bibfnamefont {C.}~\bibnamefont {Faria}}, \bibinfo {author} {\bibfnamefont {I.}~\bibnamefont {Chagas}},\ and\ \bibinfo {author} {\bibfnamefont {C.}~\bibnamefont {Galvāo}},\ }\bibfield  {title} {\bibinfo {title} {Promoting {Science} {Outdoor} {Activities} for {Elementary} {School} {Children}: {Contributions} from a research laboratory},\ }\href {https://doi.org/10.1080/09500693.2011.583292} {\bibfield  {journal} {\bibinfo  {journal} {International Journal of Science Education}\ }\textbf {\bibinfo {volume} {35}},\ \bibinfo {pages} {796} (\bibinfo {year} {2013})}\BibitemShut {NoStop}%
\bibitem [{\citenamefont {Kuhn}\ \emph {et~al.}(1995)\citenamefont {Kuhn}, \citenamefont {Garcia-Mila}, \citenamefont {Zohar}, \citenamefont {Andersen}, \citenamefont {White}, \citenamefont {Klahr},\ and\ \citenamefont {Carver}}]{Kuhn1995}%
  \BibitemOpen
  \bibfield  {author} {\bibinfo {author} {\bibfnamefont {D.}~\bibnamefont {Kuhn}}, \bibinfo {author} {\bibfnamefont {M.}~\bibnamefont {Garcia-Mila}}, \bibinfo {author} {\bibfnamefont {A.}~\bibnamefont {Zohar}}, \bibinfo {author} {\bibfnamefont {C.}~\bibnamefont {Andersen}}, \bibinfo {author} {\bibfnamefont {S.~H.}\ \bibnamefont {White}}, \bibinfo {author} {\bibfnamefont {D.}~\bibnamefont {Klahr}},\ and\ \bibinfo {author} {\bibfnamefont {S.~M.}\ \bibnamefont {Carver}},\ }\bibfield  {title} {\bibinfo {title} {Strategies of {Knowledge} {Acquisition}},\ }\bibfield  {journal} {\bibinfo  {journal} {Monographs of the society for research in child development}\ }\textbf {\bibinfo {volume} {60}},\ \href {https://doi.org/10.2307/1166059} {10.2307/1166059} (\bibinfo {year} {1995})\BibitemShut {NoStop}%
\bibitem [{\citenamefont {Kanari}\ and\ \citenamefont {Millar}(2004)}]{Kanari2004}%
  \BibitemOpen
  \bibfield  {author} {\bibinfo {author} {\bibfnamefont {Z.}~\bibnamefont {Kanari}}\ and\ \bibinfo {author} {\bibfnamefont {R.}~\bibnamefont {Millar}},\ }\bibfield  {title} {\bibinfo {title} {Reasoning from data: {How} students collect and interpret data in science investigations},\ }\href {https://doi.org/10.1002/tea.20020} {\bibfield  {journal} {\bibinfo  {journal} {Journal of Research in Science Teaching}\ }\textbf {\bibinfo {volume} {41}},\ \bibinfo {pages} {748} (\bibinfo {year} {2004})}\BibitemShut {NoStop}%
\bibitem [{\citenamefont {Baur}(2018)}]{Baur2018}%
  \BibitemOpen
  \bibfield  {author} {\bibinfo {author} {\bibfnamefont {A.}~\bibnamefont {Baur}},\ }\bibfield  {title} {\bibinfo {title} {Fehler, {Fehlkonzepte} und spezifische {Vorgehensweisen} von {Schülerinnen} und {Schülern} beim {Experimentieren} [{Mistakes}, {Misconceptions}, and {Pupils}’ {Idiosyncratic} {Approaches} to {Experimentation} {Findings} from an {Observation}]},\ }\href {https://doi.org/10.1007/s40573-018-0078-7} {\bibfield  {journal} {\bibinfo  {journal} {Zeitschrift für Didaktik der Naturwissenschaften}\ }\textbf {\bibinfo {volume} {24}},\ \bibinfo {pages} {115} (\bibinfo {year} {2018})}\BibitemShut {NoStop}%
\bibitem [{\citenamefont {Garcia-Mila}\ \emph {et~al.}(2017)\citenamefont {Garcia-Mila}, \citenamefont {Criado},\ and\ \citenamefont {Cruz-Guzmán}}]{GarciaMila2017}%
  \BibitemOpen
  \bibfield  {author} {\bibinfo {author} {\bibfnamefont {M.}~\bibnamefont {Garcia-Mila}}, \bibinfo {author} {\bibfnamefont {A.~M.}\ \bibnamefont {Criado}},\ and\ \bibinfo {author} {\bibfnamefont {M.}~\bibnamefont {Cruz-Guzmán}},\ }\bibfield  {title} {\bibinfo {title} {Primary pre-service teachers’ skills in planning a guided scientific inquiry},\ }\href {https://doi.org/10.1007/s11165-016-9536-8} {\bibfield  {journal} {\bibinfo  {journal} {Research in Science Education}\ }\textbf {\bibinfo {volume} {47}},\ \bibinfo {pages} {989} (\bibinfo {year} {2017})}\BibitemShut {NoStop}%
\bibitem [{\citenamefont {Ludwig}\ \emph {et~al.}(2021)\citenamefont {Ludwig}, \citenamefont {Priemer},\ and\ \citenamefont {Lewalter}}]{Ludwig2021}%
  \BibitemOpen
  \bibfield  {author} {\bibinfo {author} {\bibfnamefont {T.}~\bibnamefont {Ludwig}}, \bibinfo {author} {\bibfnamefont {B.}~\bibnamefont {Priemer}},\ and\ \bibinfo {author} {\bibfnamefont {D.}~\bibnamefont {Lewalter}},\ }\bibfield  {title} {\bibinfo {title} {Assessing {Secondary} {School} {Students}' {Justifications} for {Supporting} or {Rejecting} a {Scientific} {Hypothesis} in the {Physics} {Lab}},\ }\href {https://doi.org/10.1007/s11165-019-09862-4} {\bibfield  {journal} {\bibinfo  {journal} {Research in Science Education}\ }\textbf {\bibinfo {volume} {51}},\ \bibinfo {pages} {819} (\bibinfo {year} {2021})}\BibitemShut {NoStop}%
\bibitem [{\citenamefont {Lenz}(2025)}]{Lenz2025}%
  \BibitemOpen
  \bibfield  {author} {\bibinfo {author} {\bibfnamefont {L.}~\bibnamefont {Lenz}},\ }\emph {\bibinfo {title} {Hindernisse beim datenbasierten {Begründen} physikalischer {Hypothesen} in {Experimentiersettings} [Challenges in data-driven justification of physical hypotheses in experimental settings]}},\ \href {urn:nbn:de:bsz:751-opus4-6989} {\bibinfo {type} {Dissertation}},\ \bibinfo  {school} {Karlsruhe University of Education}, \bibinfo {address} {Karlsruhe} (\bibinfo {year} {2025})\BibitemShut {NoStop}%
\bibitem [{\citenamefont {Greenhoot}\ \emph {et~al.}(2004)\citenamefont {Greenhoot}, \citenamefont {Semb}, \citenamefont {Colombo},\ and\ \citenamefont {Schreiber}}]{Greenhoot2004}%
  \BibitemOpen
  \bibfield  {author} {\bibinfo {author} {\bibfnamefont {A.~F.}\ \bibnamefont {Greenhoot}}, \bibinfo {author} {\bibfnamefont {G.}~\bibnamefont {Semb}}, \bibinfo {author} {\bibfnamefont {J.}~\bibnamefont {Colombo}},\ and\ \bibinfo {author} {\bibfnamefont {T.}~\bibnamefont {Schreiber}},\ }\bibfield  {title} {\bibinfo {title} {Prior beliefs and methodological concepts in scientific reasoning},\ }\href {https://doi.org/10.1002/acp.959} {\bibfield  {journal} {\bibinfo  {journal} {Applied Cognitive Psychology}\ }\textbf {\bibinfo {volume} {18}},\ \bibinfo {pages} {203} (\bibinfo {year} {2004})}\BibitemShut {NoStop}%
\bibitem [{\citenamefont {Valanides}\ \emph {et~al.}(2014)\citenamefont {Valanides}, \citenamefont {Papageorgiou},\ and\ \citenamefont {Angeli}}]{Valanides2014}%
  \BibitemOpen
  \bibfield  {author} {\bibinfo {author} {\bibfnamefont {N.}~\bibnamefont {Valanides}}, \bibinfo {author} {\bibfnamefont {M.}~\bibnamefont {Papageorgiou}},\ and\ \bibinfo {author} {\bibfnamefont {C.}~\bibnamefont {Angeli}},\ }\bibfield  {title} {\bibinfo {title} {Scientific {Investigations} of {Elementary} {School} {Children}},\ }\href {https://doi.org/10.1007/s10956-013-9448-6} {\bibfield  {journal} {\bibinfo  {journal} {Journal of Science Education and Technology}\ }\textbf {\bibinfo {volume} {23}},\ \bibinfo {pages} {26} (\bibinfo {year} {2014})}\BibitemShut {NoStop}%
\bibitem [{\citenamefont {Dunbar}(1993)}]{Dunbar1993}%
  \BibitemOpen
  \bibfield  {author} {\bibinfo {author} {\bibfnamefont {K.}~\bibnamefont {Dunbar}},\ }\bibfield  {title} {\bibinfo {title} {Concept {Discovery} in a {Scientific} {Domain}},\ }\href {https://doi.org/10.1207/s15516709cog1703_3} {\bibfield  {journal} {\bibinfo  {journal} {Cognitive Science: A Multidisciplinary Journal}\ }\textbf {\bibinfo {volume} {17}},\ \bibinfo {pages} {387} (\bibinfo {year} {1993})}\BibitemShut {NoStop}%
\bibitem [{\citenamefont {Hammann}\ \emph {et~al.}(2008)\citenamefont {Hammann}, \citenamefont {Phan}, \citenamefont {Ehmer},\ and\ \citenamefont {Grimm}}]{Hammann2008}%
  \BibitemOpen
  \bibfield  {author} {\bibinfo {author} {\bibfnamefont {M.}~\bibnamefont {Hammann}}, \bibinfo {author} {\bibfnamefont {T.~T.~H.}\ \bibnamefont {Phan}}, \bibinfo {author} {\bibfnamefont {M.}~\bibnamefont {Ehmer}},\ and\ \bibinfo {author} {\bibfnamefont {T.}~\bibnamefont {Grimm}},\ }\bibfield  {title} {\bibinfo {title} {Assessing pupils' skills in experimentation},\ }\href {https://doi.org/10.1080/00219266.2008.9656113} {\bibfield  {journal} {\bibinfo  {journal} {Journal of Biological Education}\ }\textbf {\bibinfo {volume} {42}},\ \bibinfo {pages} {66} (\bibinfo {year} {2008})}\BibitemShut {NoStop}%
\bibitem [{\citenamefont {Park}(2006)}]{Park2006}%
  \BibitemOpen
  \bibfield  {author} {\bibinfo {author} {\bibfnamefont {J.}~\bibnamefont {Park}},\ }\bibfield  {title} {\bibinfo {title} {Modelling {Analysis} of {Students}’ {Processes} of {Generating} {Scientific} {Explanatory} {Hypotheses}},\ }\href {https://doi.org/10.1080/09500690500404540} {\bibfield  {journal} {\bibinfo  {journal} {International Journal of Science Education}\ }\textbf {\bibinfo {volume} {28}},\ \bibinfo {pages} {469} (\bibinfo {year} {2006})}\BibitemShut {NoStop}%
\bibitem [{\citenamefont {Ludwig}(2017)}]{Ludwig2017}%
  \BibitemOpen
  \bibfield  {author} {\bibinfo {author} {\bibfnamefont {T.}~\bibnamefont {Ludwig}},\ }\emph {\bibinfo {title} {Argumentieren beim {Experimentieren} - {Die} {Bedeutung} personaler und situationaler {Faktoren} [{Argumentation} in {Inquiry}—{The} influence of {Personal} and {Situational} {Factors}]}},\ \href {10.18452/18408} {\bibinfo {type} {Dissertation}},\ \bibinfo  {school} {Humboldt Universität zu Berlin}, \bibinfo {address} {Berlin} (\bibinfo {year} {2017})\BibitemShut {NoStop}%
\bibitem [{\citenamefont {Hahn}\ and\ \citenamefont {Klein}(2022)}]{Hahn2022}%
  \BibitemOpen
  \bibfield  {author} {\bibinfo {author} {\bibfnamefont {L.}~\bibnamefont {Hahn}}\ and\ \bibinfo {author} {\bibfnamefont {P.}~\bibnamefont {Klein}},\ }\bibfield  {title} {\bibinfo {title} {Eye tracking in physics education research: {A} systematic literature review},\ }\href {https://doi.org/10.1103/PhysRevPhysEducRes.18.013102} {\bibfield  {journal} {\bibinfo  {journal} {Physical Review Physics Education Research}\ }\textbf {\bibinfo {volume} {18}},\ \bibinfo {pages} {013102} (\bibinfo {year} {2022})}\BibitemShut {NoStop}%
\bibitem [{\citenamefont {Becker}\ \emph {et~al.}(2023)\citenamefont {Becker}, \citenamefont {Knippertz}, \citenamefont {Ruzika},\ and\ \citenamefont {Kuhn}}]{Becker2023}%
  \BibitemOpen
  \bibfield  {author} {\bibinfo {author} {\bibfnamefont {S.}~\bibnamefont {Becker}}, \bibinfo {author} {\bibfnamefont {L.}~\bibnamefont {Knippertz}}, \bibinfo {author} {\bibfnamefont {S.}~\bibnamefont {Ruzika}},\ and\ \bibinfo {author} {\bibfnamefont {J.}~\bibnamefont {Kuhn}},\ }\bibfield  {title} {\bibinfo {title} {Persistence, context, and visual strategy of graph understanding: {Gaze} patterns reveal student difficulties in interpreting graphs},\ }\href {https://doi.org/10.1103/PhysRevPhysEducRes.19.020142} {\bibfield  {journal} {\bibinfo  {journal} {Physical Review Physics Education Research}\ }\textbf {\bibinfo {volume} {19}},\ \bibinfo {pages} {020142} (\bibinfo {year} {2023})}\BibitemShut {NoStop}%
\bibitem [{\citenamefont {Just}\ and\ \citenamefont {Carpenter}(1980)}]{Carpenter1989}%
  \BibitemOpen
  \bibfield  {author} {\bibinfo {author} {\bibfnamefont {M.~A.}\ \bibnamefont {Just}}\ and\ \bibinfo {author} {\bibfnamefont {S.}~\bibnamefont {Carpenter}},\ }\bibfield  {title} {\bibinfo {title} {A theory of reading: {From} eye fixations to comprehension},\ }\href {https://doi.org/10.1037/0033-295X.87.4.329} {\bibfield  {journal} {\bibinfo  {journal} {Psychological Review}\ }\textbf {\bibinfo {volume} {87}},\ \bibinfo {pages} {329} (\bibinfo {year} {1980})}\BibitemShut {NoStop}%
\bibitem [{\citenamefont {Madsen}\ \emph {et~al.}(2012)\citenamefont {Madsen}, \citenamefont {Larson}, \citenamefont {Loschky},\ and\ \citenamefont {Rebello}}]{Madsen2012}%
  \BibitemOpen
  \bibfield  {author} {\bibinfo {author} {\bibfnamefont {A.~M.}\ \bibnamefont {Madsen}}, \bibinfo {author} {\bibfnamefont {A.~M.}\ \bibnamefont {Larson}}, \bibinfo {author} {\bibfnamefont {L.~C.}\ \bibnamefont {Loschky}},\ and\ \bibinfo {author} {\bibfnamefont {N.~S.}\ \bibnamefont {Rebello}},\ }\bibfield  {title} {\bibinfo {title} {Differences in visual attention between those who correctly and incorrectly answer physics problems},\ }\href {https://doi.org/10.1103/PhysRevSTPER.8.010122} {\bibfield  {journal} {\bibinfo  {journal} {Physical Review Physics Education Research}\ }\textbf {\bibinfo {volume} {8}},\ \bibinfo {pages} {010122} (\bibinfo {year} {2012})}\BibitemShut {NoStop}%
\bibitem [{\citenamefont {Madsen}\ \emph {et~al.}(2013)\citenamefont {Madsen}, \citenamefont {Rouinfar}, \citenamefont {Larson}, \citenamefont {Loschky},\ and\ \citenamefont {Rebello}}]{Madsen2013}%
  \BibitemOpen
  \bibfield  {author} {\bibinfo {author} {\bibfnamefont {A.~M.}\ \bibnamefont {Madsen}}, \bibinfo {author} {\bibfnamefont {A.}~\bibnamefont {Rouinfar}}, \bibinfo {author} {\bibfnamefont {A.~M.}\ \bibnamefont {Larson}}, \bibinfo {author} {\bibfnamefont {L.~C.}\ \bibnamefont {Loschky}},\ and\ \bibinfo {author} {\bibfnamefont {N.~S.}\ \bibnamefont {Rebello}},\ }\bibfield  {title} {\bibinfo {title} {Can short duration visual cues influence students' reasoning and eye movements in physics problems?},\ }\href {https://doi.org/10.1103/PhysRevSTPER.9.020104} {\bibfield  {journal} {\bibinfo  {journal} {Physical Review Physics Education Research}\ }\textbf {\bibinfo {volume} {9}},\ \bibinfo {pages} {020104} (\bibinfo {year} {2013})}\BibitemShut {NoStop}%
\bibitem [{\citenamefont {Ruf}\ \emph {et~al.}(2023)\citenamefont {Ruf}, \citenamefont {Horrer}, \citenamefont {Berndt}, \citenamefont {Hofer}, \citenamefont {Fischer}, \citenamefont {Zottmann}, \citenamefont {Kuhn},\ and\ \citenamefont {Küchemann}}]{Ruf2023}%
  \BibitemOpen
  \bibfield  {author} {\bibinfo {author} {\bibfnamefont {V.}~\bibnamefont {Ruf}}, \bibinfo {author} {\bibfnamefont {A.}~\bibnamefont {Horrer}}, \bibinfo {author} {\bibfnamefont {M.}~\bibnamefont {Berndt}}, \bibinfo {author} {\bibfnamefont {S.}~\bibnamefont {Hofer}}, \bibinfo {author} {\bibfnamefont {F.}~\bibnamefont {Fischer}}, \bibinfo {author} {\bibfnamefont {J.}~\bibnamefont {Zottmann}}, \bibinfo {author} {\bibfnamefont {J.}~\bibnamefont {Kuhn}},\ and\ \bibinfo {author} {\bibfnamefont {S.}~\bibnamefont {Küchemann}},\ }\bibfield  {title} {\bibinfo {title} {A {Literature} {Review} {Comparing} {Experts}' and {Non}-{Experts}' {Visual} {Processing} of {Graphs} during {Problem}-{Solving} and {Learning}},\ }\href {https://doi.org/10.3390/educsci13020216} {\bibfield  {journal} {\bibinfo  {journal} {Education Sciences}\ }\textbf {\bibinfo {volume} {13}},\ \bibinfo {pages} {216} (\bibinfo {year} {2023})}\BibitemShut {NoStop}%
\bibitem [{\citenamefont {Brückner}\ \emph {et~al.}(2020{\natexlab{a}})\citenamefont {Brückner}, \citenamefont {Schneider}, \citenamefont {Zlatkin-Troitschanskaia},\ and\ \citenamefont {Drachsler}}]{Brückner2020}%
  \BibitemOpen
  \bibfield  {author} {\bibinfo {author} {\bibfnamefont {S.}~\bibnamefont {Brückner}}, \bibinfo {author} {\bibfnamefont {J.}~\bibnamefont {Schneider}}, \bibinfo {author} {\bibfnamefont {O.}~\bibnamefont {Zlatkin-Troitschanskaia}},\ and\ \bibinfo {author} {\bibfnamefont {H.}~\bibnamefont {Drachsler}},\ }\bibfield  {title} {\bibinfo {title} {Epistemic {Network} {Analyses} of {Economics} {Students}' {Graph} {Understanding}: {An} {Eye}-{Tracking} {Study}},\ }\href {https://doi.org/10.3390/s20236908} {\bibfield  {journal} {\bibinfo  {journal} {Sensors}\ }\textbf {\bibinfo {volume} {20}},\ \bibinfo {pages} {6908} (\bibinfo {year} {2020}{\natexlab{a}})}\BibitemShut {NoStop}%
\bibitem [{\citenamefont {Brückner}\ \emph {et~al.}(2020{\natexlab{b}})\citenamefont {Brückner}, \citenamefont {Zlatkin-Troitschanskaia}, \citenamefont {Küchemann}, \citenamefont {Klein},\ and\ \citenamefont {Kuhn}}]{Brueckner2020}%
  \BibitemOpen
  \bibfield  {author} {\bibinfo {author} {\bibfnamefont {S.}~\bibnamefont {Brückner}}, \bibinfo {author} {\bibfnamefont {O.}~\bibnamefont {Zlatkin-Troitschanskaia}}, \bibinfo {author} {\bibfnamefont {S.}~\bibnamefont {Küchemann}}, \bibinfo {author} {\bibfnamefont {P.}~\bibnamefont {Klein}},\ and\ \bibinfo {author} {\bibfnamefont {J.}~\bibnamefont {Kuhn}},\ }\bibfield  {title} {\bibinfo {title} {Changes in {Students}' {Understanding} of and {Visual} {Attention} on {Digitally} {Represented} {Graphs} {Across} {Two} {Domains} in {Higher} {Education}: {A} {Postreplication} {Study}},\ }\href {https://doi.org/10.3389/fpsyg.2020.02090} {\bibfield  {journal} {\bibinfo  {journal} {Fontiers in Psychology}\ }\textbf {\bibinfo {volume} {11}},\ \bibinfo {pages} {2090} (\bibinfo {year} {2020}{\natexlab{b}})}\BibitemShut {NoStop}%
\bibitem [{\citenamefont {Kekule}(2014)}]{Kekule2014}%
  \BibitemOpen
  \bibfield  {author} {\bibinfo {author} {\bibfnamefont {M.}~\bibnamefont {Kekule}},\ }\bibfield  {title} {\bibinfo {title} {Students' approaches when dealing with kinematics graphs explored by eye-tracking research method},\ }\href {https://doi.org/10.30935/scimath/9632} {\bibfield  {journal} {\bibinfo  {journal} {European Journal of Science and Mathematics Education}\ }\textbf {\bibinfo {volume} {2}},\ \bibinfo {pages} {108} (\bibinfo {year} {2014})}\BibitemShut {NoStop}%
\bibitem [{\citenamefont {Klein}\ \emph {et~al.}(2019)\citenamefont {Klein}, \citenamefont {Küchemann}, \citenamefont {Brückner}, \citenamefont {Zlatkin-Troitschanskaia},\ and\ \citenamefont {Kuhn}}]{Klein2019}%
  \BibitemOpen
  \bibfield  {author} {\bibinfo {author} {\bibfnamefont {P.}~\bibnamefont {Klein}}, \bibinfo {author} {\bibfnamefont {S.}~\bibnamefont {Küchemann}}, \bibinfo {author} {\bibfnamefont {S.}~\bibnamefont {Brückner}}, \bibinfo {author} {\bibfnamefont {O.}~\bibnamefont {Zlatkin-Troitschanskaia}},\ and\ \bibinfo {author} {\bibfnamefont {J.}~\bibnamefont {Kuhn}},\ }\bibfield  {title} {\bibinfo {title} {Student understanding of graph slope and area under a curve: {A} replication study comparing first-year physics and economics students},\ }\href {https://doi.org/10.1103/PhysRevPhysEducRes.15.020116} {\bibfield  {journal} {\bibinfo  {journal} {Physical Review Physics Education Research}\ }\textbf {\bibinfo {volume} {15}},\ \bibinfo {pages} {020116} (\bibinfo {year} {2019})}\BibitemShut {NoStop}%
\bibitem [{\citenamefont {Klein}\ \emph {et~al.}(2020)\citenamefont {Klein}, \citenamefont {Lichtenberger}, \citenamefont {Küchemann}, \citenamefont {Becker}, \citenamefont {Kekule}, \citenamefont {Viiri}, \citenamefont {Baadte}, \citenamefont {Vaterlaus},\ and\ \citenamefont {Kuhn}}]{Klein2020}%
  \BibitemOpen
  \bibfield  {author} {\bibinfo {author} {\bibfnamefont {P.}~\bibnamefont {Klein}}, \bibinfo {author} {\bibfnamefont {A.}~\bibnamefont {Lichtenberger}}, \bibinfo {author} {\bibfnamefont {S.}~\bibnamefont {Küchemann}}, \bibinfo {author} {\bibfnamefont {S.}~\bibnamefont {Becker}}, \bibinfo {author} {\bibfnamefont {M.}~\bibnamefont {Kekule}}, \bibinfo {author} {\bibfnamefont {J.}~\bibnamefont {Viiri}}, \bibinfo {author} {\bibfnamefont {C.}~\bibnamefont {Baadte}}, \bibinfo {author} {\bibfnamefont {A.}~\bibnamefont {Vaterlaus}},\ and\ \bibinfo {author} {\bibfnamefont {J.}~\bibnamefont {Kuhn}},\ }\bibfield  {title} {\bibinfo {title} {Visual attention while solving the test of understanding graphs in kinematics: an eye-tracking analysis},\ }\href {https://doi.org/10.1088/1361-6404/ab5f51} {\bibfield  {journal} {\bibinfo  {journal} {European Journal of Physics}\ }\textbf {\bibinfo {volume} {41}},\ \bibinfo {pages} {025701} (\bibinfo {year} {2020})}\BibitemShut {NoStop}%
\bibitem [{\citenamefont {Klein}\ \emph {et~al.}(2021)\citenamefont {Klein}, \citenamefont {Becker}, \citenamefont {Küchemann},\ and\ \citenamefont {Kuhn}}]{Klein2021}%
  \BibitemOpen
  \bibfield  {author} {\bibinfo {author} {\bibfnamefont {P.}~\bibnamefont {Klein}}, \bibinfo {author} {\bibfnamefont {S.}~\bibnamefont {Becker}}, \bibinfo {author} {\bibfnamefont {S.}~\bibnamefont {Küchemann}},\ and\ \bibinfo {author} {\bibfnamefont {J.}~\bibnamefont {Kuhn}},\ }\bibfield  {title} {\bibinfo {title} {Test of understanding graphs in kinematics: {Item} objectives confirmed by clustering eye movement transitions},\ }\href {https://doi.org/10.1103/PhysRevPhysEducRes.17.013102} {\bibfield  {journal} {\bibinfo  {journal} {Physical Review Physics Education Research}\ }\textbf {\bibinfo {volume} {17}},\ \bibinfo {pages} {013102} (\bibinfo {year} {2021})}\BibitemShut {NoStop}%
\bibitem [{\citenamefont {Rouinfar}\ \emph {et~al.}(2014)\citenamefont {Rouinfar}, \citenamefont {Agra}, \citenamefont {Larson}, \citenamefont {Rebello},\ and\ \citenamefont {Loschky}}]{Rouinfar2014}%
  \BibitemOpen
  \bibfield  {author} {\bibinfo {author} {\bibfnamefont {A.}~\bibnamefont {Rouinfar}}, \bibinfo {author} {\bibfnamefont {E.}~\bibnamefont {Agra}}, \bibinfo {author} {\bibfnamefont {A.~M.}\ \bibnamefont {Larson}}, \bibinfo {author} {\bibfnamefont {N.~S.}\ \bibnamefont {Rebello}},\ and\ \bibinfo {author} {\bibfnamefont {L.~C.}\ \bibnamefont {Loschky}},\ }\bibfield  {title} {\bibinfo {title} {Linking attentional processes and conceptual problem solving: visual cues facilitate the automaticity of extracting relevant information from diagrams},\ }\href {https://doi.org/10.3389/fpsyg.2014.01094} {\bibfield  {journal} {\bibinfo  {journal} {Fontiers in Psychology}\ }\textbf {\bibinfo {volume} {5}},\ \bibinfo {pages} {1094} (\bibinfo {year} {2014})}\BibitemShut {NoStop}%
\bibitem [{\citenamefont {Skrabankova}\ \emph {et~al.}(2020)\citenamefont {Skrabankova}, \citenamefont {Popelka},\ and\ \citenamefont {Beitlova}}]{Skrabankova2020}%
  \BibitemOpen
  \bibfield  {author} {\bibinfo {author} {\bibfnamefont {J.}~\bibnamefont {Skrabankova}}, \bibinfo {author} {\bibfnamefont {S.}~\bibnamefont {Popelka}},\ and\ \bibinfo {author} {\bibfnamefont {M.}~\bibnamefont {Beitlova}},\ }\bibfield  {title} {\bibinfo {title} {Students' ability to work with graphs in physics studies related to three typical student groups},\ }\href {https://doi.org/10.33225/jbse/20.19.298} {\bibfield  {journal} {\bibinfo  {journal} {Journal of Baltic Science Education}\ }\textbf {\bibinfo {volume} {19}},\ \bibinfo {pages} {298} (\bibinfo {year} {2020})}\BibitemShut {NoStop}%
\bibitem [{\citenamefont {Susac}\ \emph {et~al.}(2018)\citenamefont {Susac}, \citenamefont {Bubic}, \citenamefont {Kazotti}, \citenamefont {Planinic},\ and\ \citenamefont {Palmovic}}]{Susac2018}%
  \BibitemOpen
  \bibfield  {author} {\bibinfo {author} {\bibfnamefont {A.}~\bibnamefont {Susac}}, \bibinfo {author} {\bibfnamefont {A.}~\bibnamefont {Bubic}}, \bibinfo {author} {\bibfnamefont {E.}~\bibnamefont {Kazotti}}, \bibinfo {author} {\bibfnamefont {M.}~\bibnamefont {Planinic}},\ and\ \bibinfo {author} {\bibfnamefont {M.}~\bibnamefont {Palmovic}},\ }\bibfield  {title} {\bibinfo {title} {Student understanding of graph slope and area under a graph: {A} comparison of physics and nonphysics students},\ }\href {https://doi.org/10.1103/PhysRevPhysEducRes.14.020109} {\bibfield  {journal} {\bibinfo  {journal} {Physical Review Physics Education Research}\ }\textbf {\bibinfo {volume} {14}},\ \bibinfo {pages} {020109} (\bibinfo {year} {2018})}\BibitemShut {NoStop}%
\bibitem [{\citenamefont {Kok}\ and\ \citenamefont {Jarodzka}(2016)}]{Kok2016}%
  \BibitemOpen
  \bibfield  {author} {\bibinfo {author} {\bibfnamefont {E.~M.}\ \bibnamefont {Kok}}\ and\ \bibinfo {author} {\bibfnamefont {H.}~\bibnamefont {Jarodzka}},\ }\bibfield  {title} {\bibinfo {title} {Before your very eyes: the value and limitations of eye tracking in medical education},\ }\href {https://doi.org/10.1111/medu.13066} {\bibfield  {journal} {\bibinfo  {journal} {Medical Education}\ }\textbf {\bibinfo {volume} {51}},\ \bibinfo {pages} {114} (\bibinfo {year} {2016})}\BibitemShut {NoStop}%
\bibitem [{\citenamefont {Klassen}\ and\ \citenamefont {Friege}(2026)}]{Klassen2026}%
  \BibitemOpen
  \bibfield  {author} {\bibinfo {author} {\bibfnamefont {M.}~\bibnamefont {Klassen}}\ and\ \bibinfo {author} {\bibfnamefont {G.}~\bibnamefont {Friege}},\ }\bibfield  {title} {\bibinfo {title} {Difficulties in diagram interpretation in physics: {An} eye-tracking study of visual attention and diagram interpretation strategies},\ }\href {https://doi.org/10.1103/58hn-493c} {\bibfield  {journal} {\bibinfo  {journal} {Physical Review Physics Education Research}\ }\textbf {\bibinfo {volume} {22}},\ \bibinfo {pages} {010121} (\bibinfo {year} {2026})}\BibitemShut {NoStop}%
\bibitem [{\citenamefont {Kiili}\ \emph {et~al.}(2014)\citenamefont {Kiili}, \citenamefont {Ketamo}, \citenamefont {Koivisto},\ and\ \citenamefont {Finn}}]{Kiili2014}%
  \BibitemOpen
  \bibfield  {author} {\bibinfo {author} {\bibfnamefont {K.}~\bibnamefont {Kiili}}, \bibinfo {author} {\bibfnamefont {H.}~\bibnamefont {Ketamo}}, \bibinfo {author} {\bibfnamefont {A.}~\bibnamefont {Koivisto}},\ and\ \bibinfo {author} {\bibfnamefont {E.}~\bibnamefont {Finn}},\ }\bibfield  {title} {\bibinfo {title} {Studying the {User} {Experience} of a {Tablet} {Based} {Math} {Game}},\ }\href {https://doi.org/10.4018/IJGBL.2014010104} {\bibfield  {journal} {\bibinfo  {journal} {International Journal of Game-Based Learning}\ }\textbf {\bibinfo {volume} {4}},\ \bibinfo {pages} {60} (\bibinfo {year} {2014})}\BibitemShut {NoStop}%
\bibitem [{\citenamefont {Ibrahim}\ and\ \citenamefont {Ding}(2021)}]{Ibrahim2021}%
  \BibitemOpen
  \bibfield  {author} {\bibinfo {author} {\bibfnamefont {B.}~\bibnamefont {Ibrahim}}\ and\ \bibinfo {author} {\bibfnamefont {L.}~\bibnamefont {Ding}},\ }\bibfield  {title} {\bibinfo {title} {Sequential and simultaneous synthesis problem solving: {A} comparison of students' gaze transitions},\ }\href {https://doi.org/10.1103/PhysRevPhysEducRes.17.010126} {\bibfield  {journal} {\bibinfo  {journal} {Physical Review Physics Education Research}\ }\textbf {\bibinfo {volume} {17}},\ \bibinfo {pages} {010126} (\bibinfo {year} {2021})}\BibitemShut {NoStop}%
\bibitem [{\citenamefont {AB}(2024)}]{TobiSoftware2024}%
  \BibitemOpen
  \bibfield  {author} {\bibinfo {author} {\bibfnamefont {T.}~\bibnamefont {AB}},\ }\href {https://www.tobii.com/} {\bibinfo {title} {Tobii {Pro} {Lab}}} (\bibinfo {year} {2024}),\ \bibinfo {note} {\url{https://www.tobii.com/}}\BibitemShut {NoStop}%
\bibitem [{\citenamefont {Salvucci}\ and\ \citenamefont {Goldberg}(2000)}]{Salvucci2000}%
  \BibitemOpen
  \bibfield  {author} {\bibinfo {author} {\bibfnamefont {D.~D.}\ \bibnamefont {Salvucci}}\ and\ \bibinfo {author} {\bibfnamefont {J.~H.}\ \bibnamefont {Goldberg}},\ }\bibfield  {title} {\bibinfo {title} {Identifying {Fixations} and {Saccades} in {Eye}-{Tracking} {Protocols}},\ }in\ \href@noop {} {\emph {\bibinfo {booktitle} {Proceedings of the {Eye} {Tracking} {Research} and {Applications} {Symposium}}}}\ (\bibinfo  {publisher} {ACM Press},\ \bibinfo {address} {New York},\ \bibinfo {year} {2000})\ pp.\ \bibinfo {pages} {71--78}\BibitemShut {NoStop}%
\bibitem [{\citenamefont {Landis}\ and\ \citenamefont {Koch}(1977)}]{Landis1977}%
  \BibitemOpen
  \bibfield  {author} {\bibinfo {author} {\bibfnamefont {J.~R.}\ \bibnamefont {Landis}}\ and\ \bibinfo {author} {\bibfnamefont {G.~G.}\ \bibnamefont {Koch}},\ }\bibfield  {title} {\bibinfo {title} {An {Application} of {Hierarchical} {Kappa}-type {Statistics} in the {Assessment} of {Majority} {Agreement} among {Multiple} {Observers}},\ }\href {https://doi.org/10.2307/2529786} {\bibfield  {journal} {\bibinfo  {journal} {Biometrics}\ }\textbf {\bibinfo {volume} {33}},\ \bibinfo {pages} {363} (\bibinfo {year} {1977})}\BibitemShut {NoStop}%
\bibitem [{\citenamefont {Masnick}\ and\ \citenamefont {Morris}(2022)}]{Masnick2022}%
  \BibitemOpen
  \bibfield  {author} {\bibinfo {author} {\bibfnamefont {A.~M.}\ \bibnamefont {Masnick}}\ and\ \bibinfo {author} {\bibfnamefont {B.~J.}\ \bibnamefont {Morris}},\ }\bibfield  {title} {\bibinfo {title} {A {Model} of {Scientific} {Data} {Reasoning}},\ }\href {https://doi.org/10.3390/educsci12020071} {\bibfield  {journal} {\bibinfo  {journal} {Education sciences}\ }\textbf {\bibinfo {volume} {12}},\ \bibinfo {pages} {71} (\bibinfo {year} {2022})}\BibitemShut {NoStop}%
\bibitem [{\citenamefont {Morris}\ \emph {et~al.}(2022)\citenamefont {Morris}, \citenamefont {Masnick},\ and\ \citenamefont {Was}}]{Morris2022}%
  \BibitemOpen
  \bibfield  {author} {\bibinfo {author} {\bibfnamefont {B.~J.}\ \bibnamefont {Morris}}, \bibinfo {author} {\bibfnamefont {A.~M.}\ \bibnamefont {Masnick}},\ and\ \bibinfo {author} {\bibfnamefont {C.}~\bibnamefont {Was}},\ }\bibfield  {title} {\bibinfo {title} {Making {Sense} of {Data}: {Identifying} {Children}'s {Strategies} for {Data} {Comparisons}},\ }\href {https://doi.org/10.1080/15248372.2022.2100395} {\bibfield  {journal} {\bibinfo  {journal} {Journal of Cognition and Development}\ }\textbf {\bibinfo {volume} {23}},\ \bibinfo {pages} {686} (\bibinfo {year} {2022})}\BibitemShut {NoStop}%
\bibitem [{\citenamefont {Kok}\ \emph {et~al.}(2024)\citenamefont {Kok}, \citenamefont {Chroszczinsky},\ and\ \citenamefont {Priemer}}]{Kok2024}%
  \BibitemOpen
  \bibfield  {author} {\bibinfo {author} {\bibfnamefont {K.}~\bibnamefont {Kok}}, \bibinfo {author} {\bibfnamefont {S.}~\bibnamefont {Chroszczinsky}},\ and\ \bibinfo {author} {\bibfnamefont {B.}~\bibnamefont {Priemer}},\ }\bibfield  {title} {\bibinfo {title} {How to evaluate students' decisions in a data comparison problem: {Correct} decision for the wrong reasons?},\ }\href {https://doi.org/10.1103/PhysRevPhysEducRes.20.010129} {\bibfield  {journal} {\bibinfo  {journal} {Physical Review Physics Education Research}\ }\textbf {\bibinfo {volume} {20}},\ \bibinfo {pages} {010129} (\bibinfo {year} {2024})}\BibitemShut {NoStop}%
\bibitem [{\citenamefont {Buffler}\ \emph {et~al.}(2001)\citenamefont {Buffler}, \citenamefont {Allie},\ and\ \citenamefont {Lubben}}]{Buffler2001}%
  \BibitemOpen
  \bibfield  {author} {\bibinfo {author} {\bibfnamefont {A.}~\bibnamefont {Buffler}}, \bibinfo {author} {\bibfnamefont {S.}~\bibnamefont {Allie}},\ and\ \bibinfo {author} {\bibfnamefont {F.}~\bibnamefont {Lubben}},\ }\bibfield  {title} {\bibinfo {title} {The development of first year physics students' ideas abut measurement in terms of point and set paradigms},\ }\href {https://doi.org/10.1080/09500690110039567} {\bibfield  {journal} {\bibinfo  {journal} {International Journal of Science Education}\ }\textbf {\bibinfo {volume} {23}},\ \bibinfo {pages} {1137} (\bibinfo {year} {2001})}\BibitemShut {NoStop}%
\bibitem [{\citenamefont {Séré}\ \emph {et~al.}(2001)\citenamefont {Séré}, \citenamefont {Fernandez-Gonzalez}, \citenamefont {Gallegos}, \citenamefont {Gonzalez-Garcia}, \citenamefont {De~Manuel}, \citenamefont {Perales},\ and\ \citenamefont {Leach}}]{Sere2001}%
  \BibitemOpen
  \bibfield  {author} {\bibinfo {author} {\bibfnamefont {M.-G.}\ \bibnamefont {Séré}}, \bibinfo {author} {\bibfnamefont {M.}~\bibnamefont {Fernandez-Gonzalez}}, \bibinfo {author} {\bibfnamefont {J.~A.}\ \bibnamefont {Gallegos}}, \bibinfo {author} {\bibfnamefont {F.}~\bibnamefont {Gonzalez-Garcia}}, \bibinfo {author} {\bibfnamefont {E.}~\bibnamefont {De~Manuel}}, \bibinfo {author} {\bibfnamefont {F.~J.}\ \bibnamefont {Perales}},\ and\ \bibinfo {author} {\bibfnamefont {J.}~\bibnamefont {Leach}},\ }\bibfield  {title} {\bibinfo {title} {Images of {Science} {Linked} to {Labwork}: {A} {Survey} of {Secondary} {School} and {University} {Students}},\ }\href {https://doi.org/10.1023/A:1013141706723} {\bibfield  {journal} {\bibinfo  {journal} {Research in Science Education}\ }\textbf {\bibinfo {volume} {31}},\ \bibinfo {pages} {399} (\bibinfo {year} {2001})}\BibitemShut {NoStop}%
\end{thebibliography}%

\clearpage
\onecolumngrid

\appendix
\section{Coding manual for written justifications and think-aloud data \label{sec:Coding manual for written justifications and think-aloud data}}

Table~\ref{tab:Coding manual for written justifications and think-aloud data} presents the coding manual used to evaluate the written justifications and think-aloud data. The context unit and coding unit for the written justifications is the entire justification written down. The context unit for the think-aloud data are all thoughts expressed during the task; the coding unit was at least half a sentence. Multiple codings were permitted.

%\begin{table}[H] % add [H] placement to break table across pages
%\begin{ruledtabular}
\begin{longtable}{p{0.07\columnwidth} p{0.15\columnwidth} p{0.38\columnwidth} p{0.38\columnwidth}}
\caption{Coding manual for written justifications and think-aloud data\label{tab:Coding manual for written justifications and think-aloud data}}\\
\hline
Code & Category & Description & Example\\
\hline
Dat & Measurement data in general & It refers to the measurement data without clarifying which measurement variable is being referred to. This category therefore includes (measurement data-based) statements that refer generally to the movement or patterns in the measurement data. This category is only coded if it is not explicitly clear (also with considering the diagrams itself) which measurement variables are being referred to. & ``I see fluctuation in the data." \linebreak ``You cannot see the phase change from the diagram." \linebreak ``Based on the movement shown in the diagram."\\

E & Experimental setting & It refers to aspects that influence the movement of the body and are related to the experimental setup (e.g., materials used in the experiment, motor). & ``Cork as surface is relatively non-homogeneous." \linebreak ``The motor does not pull constantly on the body." \linebreak ``The motor is calibrated so that it pulls just enough on the body."\vspace{20pt}\\

\multicolumn{4}{l}{\textit{Explicit reference to force data}}\\
\hline

F-Var & Variations in force & It refers to the force acting on the body varies. This variation can be either increasing or decreasing. \linebreak In addition, this category is also coded when general reference is made to fluctuations in the force data. Furthermore, it is coded here when both an increase and a decrease are referred to in a differentiated manner. If the explanation refers to both an increase and a decrease in the force exerted, this category is also coded. & ``The blue graph decreases in the time interval between 1.7\,s and 1.8\,s […] in the time interval between 1.8\,s and 1.9\,s, the force increases again." \linebreak ``The values of the blue points fluctuate greatly."\\

F-C & Constant force & It refers to the force acting on the body constantly. \linebreak This category is also coded when it is stated that the force data fluctuates but is constant on average, as this is the final conclusion drawn from the force data. & ``Constant force means that the body moves." \\

F-CMF & Comparison of force data to maximum force measurement at 0.5\,s & It refers to comparisons of the force data with the global maximum value (at 0.5\,s; approx.. 21\,N). & ``In this interval, there is a small peak in force, but it is not as large as at the beginning." \linebreak ``Relatively low value of the blue dots compared to the first maximum value."\\

F-SP & Single force data point(s) & It refers to a) individual force data points, b) ranges in the force data that are higher or lower than other data points, or c) ranges before or after them. These do not necessarily have to be local minima or maxima; a pure reference to the previous/subsequent force data point is sufficient.\linebreak In contrast to K-Var or F-C, it must be clear that individual force data points/values were perceived in each case and not a progression, trend, or pattern in the force data. \linebreak In case that the force data value are compared to the global maximum F-CMF will be coded and not this category. & ``The transition from the static to the sliding phase can also be recognized by the measured force, which is greater for a short time than immediately before and after." \linebreak ``The force required is sometimes greater and sometimes less than the static friction force." \linebreak ``The force is approximately at the same level as before."\\

F-OT & Overall trend in force data & It refers to force data that is constant on average although variations in force data are perceived. \linebreak For this category, the variations in force do not need to be explicitly mentioned; it is sufficient to refer to an average/mean value that is calculated from a large amount of force data. It must be clear that the statement does not refer to a short time interval (e.g., from 1.7\,s to 1.9\,s, as required in the task), but to a longer period of time. & \vspace{20pt}\\

\multicolumn{4}{l}{\textit{Explicit reference to position data}}\\
\hline

P-I & Increasing position & It refers to the position increasing (permanently). It does not refer to intervals in which the position remains constant. Furthermore, there is no explicit mention that the position could be constant. If it is stated that the speed is constant (with reference to the fact that there must be a change in position – without this having to be explicitly stated), this category is also coded. & ``I see a linear increase in distance as a function of time." \linebreak ``The speed of the body is constant during this time interval."\\

P-IC & Increasing and constant intervals in position data & It refers to position data that is both increasing and constant. A permanent change between the two phases does not need to be explicitly stated here. This category is also coded when referring to changes in the trend (e.g., fluctuations) in the position data. In contrast to P-I, it must be clear here that constant position sections have also been identified. & ``Distance traveled stops briefly and then increases again." \linebreak ``Meanwhile, the points between 1.7\,s and 1.8\,s remain unchanged."\\

P-SP & Single position data point(s) & It refers to individual position data points being higher, lower, or equal to data points before or after them. & ``The small outlier in the measurements for the downward path is an indication that the body got stuck." \linebreak ``There are orange dots that are at the same height."\\

P-OT & Overall trend in position data & It refers to the position increasing on average (constantly), although constant sections are perceived. & ``If I were to insert a compensating straight line, I would get a constant slope." \vspace{20pt}\\

\multicolumn{4}{l}{\textit{Explicit reference to measurement uncertainty}}\\
\hline

MU-MU & Measurement uncertainty of data & It refers to measurement uncertainties resulting form the fact that a data point is a value that corresponds to a value in a range that can be assigned to the true value, but is influenced by random and systematic uncertainties. & ``Slight measurement uncertainties are apparent when I imagine the calibration curve as a plot."\\

MU-S & Measurement uncertainty resulting from sampling rate & It refers to measurement uncertainties resulting from the sampling rate used and the time intervals between measurements. & ``The body probably has a constant speed. However, this is not easy to see due to measurement inaccuracy." \linebreak ``The measurement data collected is too inaccurate; due to the monotony, it is not possible to clearly rule out an adhesion phase." \linebreak ``Without further data points, I cannot make any statements in this regard."\\

MU-Gen & Measurement uncertainty in general & It refers to measurement uncertainties without clarifying which form of measurement uncertainty is explicitly referred to (cf. MU-MU, MU-S). & ``Due to measurement uncertainties, I cannot make any statements."\\
\hline
\end{longtable}
%\end{ruledtabular}
%\end{table}

\newpage
\section{Coding manual for eye-tracking fixation data \label{sec:Coding manual for eye-tracking fixation data}}

Table~\ref{tab:Coding manual for eye-tracking fixation data} presents the coding manual used to evaluate the eye-tracking fixation areas within the diagrams. The categories can be rated for fixations on the force as well as on the position data.

%\begin{table}[H] % add [H] placement to break table across pages
%\begin{ruledtabular}
\begin{longtable}{p{0.08\columnwidth} p{0.45\columnwidth} p{0.45\columnwidth}}
\caption{Coding manual for eye-tracking fixation data\label{tab:Coding manual for eye-tracking fixation data}}\\
\hline
Code & Category & Fixated time intervals for the corresponding measurand\\
\hline

\multicolumn{3}{l}{\textit{Single data point}}\\
\hline

P-GM & Global maximum (at 0.5\,s) & 0.48\,s, 0.5\,s\\

P-LM & Local maximum & 0.58\,s, 0.88\,s, 1.06\,s, 1.08\,s, 1.32\,s, 1.54\,s, 1.62\,s\\

P-NR & Data point outside the task-relevant time interval (1.7\,s to 1.9\,s) & 0.7\,s, 0.8\,s, 1.0\,s, 1.2\,s, 1.3\,s, 1.4\,s, 1.46\,s, 1.5\,s, 1.6\,s \\

P-R & Data point within the task-relevant time interval (1.7\,s to 1.9\,s) & 1.7\,s, 1.8\,s, 1.86\,s, 1.88\,s, 1.9\,s\vspace{20pt}\\

\multicolumn{3}{l}{\textit{Trend in the data}}\\
\hline

T-GM & Trend in time intervals around the global maximum in force data (0.5\,s) & 0.4\,s to 0.5\,s, 0.4\,s to 0.6\,s, 0.5\,s to 0.6\,s, 0.5\,s to 0.7\,s\\

T-R & Trend in the task-relevant time interval (1.7\,s to 1.9\,s) & 1.7\,s to 1.8\,s, 1.7\,s to 1.9\,s, 1.76\,s to 1.9\,s, 1.8\,s to 1.84\,s, 1.8\,s to 1.9\,s, 1.75\,s to 1.85\,s, 1.75\,s to 1.9\,s, 1.78\,s to 1.8\,s, 1.8\,s to 1.85\,s, 1.8\,s to 1.9\,s, 1.85\,s to 1.9\,s\\

T-PR & Trend primarily in the task-relevant time interval (1.7\,s to 1.9\,s) \newline Primarily includes 0.1\,s before or after the task-relevant time interval & 1.6\,s to 1.8\,s, 1.6\,s to 1.86\,s, 1.6\,s to 1.9\,s, 1.6\,s to 2.0\,s, 1.7\,s to 2.0\,s, 1.8\,s to 2.0\,s\\

T-BM & Trend before the beginning of movement (0.4\,s/0.48\,s) & 0.0\,s to 1.5\,s, 0.1\,s to 1.6\,s\\

T-BB & Trend from the beginning of movement that do not contain trends in the task-relevant time interval (1.7\,s to 1.9\,s) & 0.4\,s to 1.6\,s, 0.5\,s to 0.9\,s, 0.4\,s to 1.6\,s, 0.5\,s to 1.3\,s, 0.5\,s to 1.5\,s, 0.6\,s to 1.0\,s, 0.6\,s to 1.3\,s\\

T-NR & Trend outside the task-relevant time interval (1.7\,s to 1.9\,s) & 0.7\,s to 1.5\,s, 0.9\,s to 1.7\,s, 1.0\,s to 1.6\,s, 1.0\,s to 1.7\,s, 1.5\,s to 1.7\,s\\

T-L & Trend in a short time interval outside the task-relevant time interval (1.7\,s to 1.9\,s) & 0.6\,s to 0.8\,s, 0.8\,s to 0.9\,s, 1.0\,s to 1.2\,s, 1.1\,s to 1.3\,s, 1.2\,s to 1.3\,s, 1.4\,s to 1.5\,s, 1.5\,s to 1.6\,s, 1.5\,s to 1.7\,s, 1.6\,s to 1.7\,s, 1.9\,s to 2.0\,s\vspace{20pt}\\

\multicolumn{3}{l}{\textit{Global pattern in the data}}\\
\hline

G-O & Global pattern since the beginning of the movement & 0.3\,s to 2.0\,s, 0.4\,s to 1.8\,s, 0.4\,s to 1.9\,s, 0.4\,s to 2.0\,s, 0.5\,s to 1.8\,s, 0.5\,s to 1.9\,s, 0.5\,s to 2.0\,s, 0.6\,s to 1.9\,s, 0.6\,s to 2.0\,s\\

G-R & (Global) pattern with a focus on the task-relevant time interval (1.7\,s to 1.9\,s) & 0.7\,s to 1.9\,s, 0.7\,s to 2.0\,s, 0.8\,s to 1.9\,s, 0.8\,s to 2.0\,s, 0.9\,s to 1.9\,s, 0.9\,s to 2.0\,s, 1.0\,s to 1.9\,s, 1.0\,s to 2.0\,s, 1.1\,s to 1.9\,s, 1.1\,s to 2.0\,s, 1.2\,s to 1.9\,s, 1.2\,s to 2.0\,s, 1.3\,s to 1.9\,s, 1.3\,s to 2.0\,s, 1.4\,s to 1.9\,s, 1.4\,s to 2.0\,s, 1.5\,s to 1.9\,s, 1.5\,s to 2.0\,s\\

G-NR & (Global) pattern without a focus on the task-relevant time interval (1.7\,s to 1.9\,s) & 0.8\,s to 1.8\,s, 0.9\,s to 1.8\,s, 1.1\,s to 1.8\,s, 1.3\,s to 1.8\,s, 1.4\,s to 1.8\,s, 1.5\,s to 1.8\,s\\

\hline

\end{longtable}
%\end{ruledtabular}
%\end{table}

\end{document}